\documentclass[11pt]{article}
\usepackage{epsf}

\begin{document}


\newtheorem{theorem}{Theorem}[section]
\newtheorem{itlemma}{Lemma}[section]
\newtheorem{itdefinition}{Definition}[section]
\newtheorem{itexample}{Example}
\newtheorem{itclaim}{Claim}[section]
\newtheorem{itproposition}{Proposition}[section]
\newtheorem{itremark}{Remark}[section]
\newtheorem{itcorollary}{Corollary}[section]

\newenvironment{example}{\begin{itexample}\rm}{\end{itexample}}
\newenvironment{definition}{\begin{itdefinition}\rm}{\end{itdefinition}}
\newenvironment{lemma}{\begin{itlemma}\rm}{\end{itlemma}}
\newenvironment{corollary}{\begin{itcorollary}\rm}{\end{itcorollary}}
\newenvironment{claim}{\begin{itclaim}\rm}{\end{itclaim}}
\newenvironment{proposition}{\begin{itproposition}\rm}{\end{itproposition}}
\newenvironment{remark}{\begin{itremark}\rm}{\end{itremark}}

\newcommand{\qed}{\hfill \halmos} 
\newcommand{\mybox}{\hfill $\Box$} 

\newcommand{\comment}[1]{}
\newcommand{\halmos}{\rule{1ex}{1.4ex}}
\newenvironment{proof}{\noindent {\em Proof}.\ }{\hspace*{\fill}$\halmos$ 
\medskip}

\def\vbar{\mathchoice{\vrule height6.3ptdepth-.5ptwidth.8pt\kern-.8pt}
   {\vrule height6.3ptdepth-.5ptwidth.8pt\kern-.8pt}
   {\vrule height4.1ptdepth-.35ptwidth.6pt\kern-.6pt}
   {\vrule height3.1ptdepth-.25ptwidth.5pt\kern-.5pt}}
\def\fudge{\mathchoice{}{}{\mkern.5mu}{\mkern.8mu}}
\def\bbc#1#2{{\rm \mkern#2mu\vbar\mkern-#2mu#1}}
\def\bbb#1{{\rm I\mkern-3.5mu #1}}
\def\bba#1#2{{\rm #1\mkern-#2mu\fudge #1}}
\def\bb#1{{\count4=`#1 \advance\count4by-64 \ifcase\count4\or\bba A{11.5}\or
   \bbb B\or\bbc C{5}\or\bbb D\or\bbb E\or\bbb F \or\bbc G{5}\or\bbb H\or
   \bbb I\or\bbc J{3}\or\bbb K\or\bbb L \or\bbb M\or\bbb N\or\bbc O{5} \or
   \bbb P\or\bbc Q{5}\or\bbb R\or\bbc S{4.2}\or\bba T{10.5}\or\bbc U{5}\or
   \bba V{12}\or\bba W{16.5}\or\bba X{11}\or\bba Y{11.7}\or\bba Z{7.5}\fi}}

\def\Q{{\bb Q}}                         
\def\N{{\bb N}}                         
\def\R{{\bb R}}                         
\def\I{{\bb Z}}                         
\def\B{{\bb B}}                         

\def\rmtr{{\rm tr}}

\def\nasymptotic{{_{\stackrel{\displaystyle\longrightarrow}
{N\rightarrow\infty}}\,\, }} 
\def\masymptotic{{_{\stackrel{\displaystyle\longrightarrow}
{m\rightarrow\infty}}\,\, }} 
\def\wasymptotic{{_{\stackrel{\displaystyle\longrightarrow}
{w\rightarrow\infty}}\,\, }} 

\def\asymptext{\raisebox{.6ex}{${_{\stackrel{\displaystyle\longrightarro
w}{x\rightarrow\pm\infty}}\,\, }$}} 
\def\epsilim{{_{\textstyle{\rm lim}}\atop_{\epsilon\rightarrow 0+}\,\, }} 

\def\beqra{\begin{eqnarray}} \def\eeqra{\end{eqnarray}}
\def\beqast{\begin{eqnarray*}} \def\eeqast{\end{eqnarray*}}
\def\beq{\begin{equation}}      \def\eeq{\end{equation}}
\def\be{\begin{enumerate}}   \def\ee{\end{enumerate}}

\def\bet{\beta}
\def\gam{\gamma}
\def\Gam{\Gamma}
\def\la{\lambda}
\def\eps{\epsilon}
\def\La{\Lambda}
\def\si{\sigma}
\def\Si{\Sigma}
\def\al{\alpha}
\def\Tha{\Theta}
\def\tha{\theta}
\def\vphi{\varphi}
\def\del{\delta}
\def\Del{\Delta}
\def\ab{\alpha\beta}
\def\om{\omega}
\def\Om{\Omega}
\def\mn{\mu\nu}
\def\mun{^{\mu}{}_{\nu}}
\def\kap{\kappa}
\def\rsi{\rho\sigma}
\def\beal{\beta\alpha}

\def\til{\tilde}
\def\rta{\rightarrow}
\def\eqv{\equiv}
\def\nab{\nabla}
\def\pa{\partial}
\def\sit{\tilde\sigma}
\def\ul{\underline}
\def\indt{\parindent2.5em}
\def\nd{\noindent}
\def\var{{1\over 2\si^2}}
\def\ivar{\left(2\pi\si^2\right)}
\def\iivar{\left({1\over 2\pi\si^2}\right)}
\def\caa{{\cal A}}
\def\cb{{\cal B}}
\def\cac{{\cal C}}
\def\cd{{\cal D}}
\def\ce{{\cal E}}
\def\cf{{\cal F}}
\def\cg{{\cal G}}
\def\ch{{\cal H}}
\def\ci{{\cal I}}
\def\cj{{\cal{J}}}
\def\ck{{\cal K}}
\def\cl{{\cal L}}
\def\cm{{\cal M}}
\def\cn{{\cal N}}
\def\cO{{\cal O}}
\def\cp{{\cal P}}
\def\cq{{\cal Q}}
\def\car{{\cal R}}
\def\cs{{\cal S}}
\def\ct{{\cal{T}}}
\def\cu{{\cal{U}}}
\def\cv{{\cal{V}}}
\def\cw{{\cal{W}}}
\def\cx{{\cal{X}}}
\def\cy{{\cal{Y}}}
\def\cz{{\cal{Z}}}

\vspace*{.0in}
\begin{center}
{\large\bf Probabilistic analysis of a differential equation for linear 
programming}
\end{center}
\vspace{-.2in}
\begin{center}
{\bf Asa Ben-Hur$^{a,b}$, Joshua Feinberg$^{c,d}$, Shmuel Fishman$^{d,e}$}\\ 
{\bf \& Hava T. Siegelmann$^{f}$}
\end{center}
\begin{center}
$^{a)}${Biochemistry Department,}\\ 
{Stanford University CA 94305}\\
$^{b)}${Faculty  of Industrial Engineering and Management,}\\ 
{Technion, Haifa 32000, Israel.}\\
$^{c)}${Physics Department,}\\
{University of Haifa at Oranim, Tivon 36006, Israel}\\
$^{d)}${Physics Department,}\\
{Technion, Israel Institute of Technology, Haifa 32000, Israel}\\
$^{e)}${Institute for Theoretical Physics}\\
{University of California\\ Santa Barbara, CA 93106, USA}\\
$^{f)}${Laboratory of Bio-computation,}\\
{Department of Computer Science,}\\
{University of Massachusetts at Amherst,}\\ 
{Amherst, MA 01003}\\
\end{center}

\vspace{250pt}
e-mail addresses: asa@barnhilltechnologies.com,~ 
joshua@physics.technion.ac.il,\\ fishman@physics.technion.ac.il,~ 
hava@mit.edu\\
proposed running head: probabilistic analysis of a differential equation for 
LP\\
All correspondence should be sent to J. Feinberg. Tel. 972-4-8292041, 
Fax: 972-4-8221514
\vfill
\pagebreak

\begin{minipage}{6.1in}
{\abstract
In this paper we address the complexity of solving linear programming problems
with a set of differential equations that converge to a fixed point
that represents the optimal solution.
Assuming a probabilistic model, where the inputs are i.i.d. Gaussian variables,
we compute the distribution of the convergence rate to the attracting
fixed point.
Using the framework of Random Matrix Theory, we derive a simple expression 
for this distribution in the asymptotic limit of large problem size.
In this limit, we find the surprising result that the distribution of the 
convergence rate is a scaling function of a single variable. This scaling 
variable combines the convergence rate with the problem size (i.e., 
the number of variables and the number of constraints). 
We also estimate numerically the distribution of the computation time
to an approximate solution, which is the time required to reach a vicinity 
of the attracting fixed point. We find that it is also a scaling function.
Using the problem size dependence of the distribution functions, we derive
high probability bounds on the convergence rates and on the computation 
times to the approximate solution.}
\end{minipage}

\vspace{100pt}
\vfill
\pagebreak

\section{Introduction}

In recent years scientists have developed new approaches to computation,
some of them based on continuous time analog systems.
Analog VLSI devices, that are often described by differential equations,
have applications in the fields of
signal processing and optimization.
Many of these devices are implementations of neural networks 
\cite{Hertz,nn-optim,wang},
or the so-called neuromorphic systems \cite{mead} which are hardware devices
whose structure is directly motivated by the workings of the brain.
In addition there is an increasing number of algorithms based on 
differential equations that solve problems such as sorting \cite{brockett}, 
linear programming \cite{faybusovich} and algebraic problems such as
singular value decomposition and finding of eigenvectors (see 
\cite{helmke-moore} and references therein).
On a more theoretical level, differential equations are known to simulate
Turing machines \cite{Branickyshort}.
The standard theory of computation and computational complexity
\cite{Papadimitriou} deals with 
computation in discrete time and in a discrete configuration space, and is 
inadequate for the description of such systems.
This work may prove useful in the analysis and comparison of 
analog computational devices (see e.g. \cite{wang,realtime}).

In a recent paper we have proposed a framework of analog computation 
based on ODE's that converge exponentially to fixed points \cite{dds2}.
In such systems it is natural to consider the 
{\em attracting fixed point as the output}.
The input can be modeled in various ways.
One possible choice is the initial condition.
This is appropriate when the aim of the computation is to decide to
which attractor out of many possible ones the system flows (see \cite{SF}).
The main problem within this approach is related to initial conditions
in the vicinity of basin boundaries.
The flow in the vicinity of the boundary is slow, resulting in very
long computation times.
Here, as in \cite{dds2} the parameters on which the vector field depends are 
the input, and  the initial condition is
part of the algorithm.
This modeling is natural for optimization problems, where one wishes
to find extrema of some function $E(x)$,
e.g. by a gradient flow
$\dot{x}=\mbox{grad}E(x)$.
An instance of the optimization problem
is specified by the parameters of $E(x)$, i.e. by the parameters
of the vector field.

The basic entity in our model of analog computation is a set of ODEs
\beq
\frac{dx}{dt}=F(x),
\label{contsys}
\eeq
where $x$ is an $n$-dimensional vector, and $F$ is an $n$-dimensional 
smooth vector field, which converges exponentially to a fixed point.
Eq. (\ref{contsys}) solves a computational problem as follows:
Given an instance of the problem, the parameters of the vector field
$F$ are set, and it is started from some pre-determined initial condition.
The result of the computation is then deduced from the fixed point that 
the system approaches.

Even though the computation happens in a real configuration space, this 
model can be considered as either a model with real inputs, as for example 
the BSS model \cite{BCSS}, or as a model with integer or rational inputs, 
depending what types of values the initial conditions are given. 
In \cite{dds2} it was argued that the time complexity in a large class of 
ODEs is the physical time that is the time parameter of the system. 
The initial condition there was assumed to be integer or rational. 
In the present paper, on the other hand, we consider real inputs. More 
specifically, we will analyze the complexity 
of a flow for linear programming (LP) introduced in \cite{faybusovich}.
In the real number model the complexity of solving LP with interior
point methods is unbounded \cite{lptraub}, and a similar phenomenon
occurs for the flow we analyze here.
To obtain finite computation times one can either measure the computation
time in terms of a {\em condition number} as in \cite{renegar-condition}, or 
impose a distribution over the set of LP instances.
Many of the probabilistic models used to study the performance of
the simplex algorithm and interior point methods
assume a Gaussian distribution of the
data \cite{lpsmale,Todd-models,shamir}, and we adopt this assumption for our 
model.
Recall that the worst case bound for the simplex algorithm is exponential
whereas some of the probabilistic bounds are quadratic \cite{shamir}.

Two types of probabilistic analysis were carried out in the LP literature:
average case and ``high probability'' behavior 
\cite{Anstreicher,Ye-book,Ye-highprob}.
A high probability analysis provides a bound on the computation time that
holds with probability 1 as the problem size goes to infinity 
\cite{Ye-highprob}.
In a worst case analysis interior point methods 
generally require $\cO(\sqrt{n} |\log \epsilon|)$
iterations to compute the cost function with $\epsilon$-precision, where
$n$ is the number of variables \cite{Ye-book}.
The high probability analysis essentially sets a limit on the required 
precision and yields $\cO(\sqrt{n} \log n)$ behavior \cite{Ye-highprob}.
However, the number of iterations has to be multiplied by the complexity
of each iteration which is $\cO(n^3)$, resulting in an overall complexity
$\cO(n^{3.5} \log n)$ in the high probability model \cite{Ye-book}.
The same factor per iteration appears in the average case analysis as well
\cite{Anstreicher}.

In contrast, in our model of analog computation, 
the computation time is the physical time required by
a hardware implementation of the vector field F(x) to converge to the
attracting fixed point. We need neither to follow the flow 
step-wise nor to calculate the vector field F(x)
since it is assumed to be realized in hardware and does not require
repetitive digital approximations. As a result, the complexity of analog 
processes does not include the $\cO (n^3)$ term as above, and in particular 
it is lower than the digital complexity of interior
point methods. In this set-up we conjecture, based on numerical calculations, 
that the flow analyzed in this paper has complexity $\cO(n\log n)$ on 
average and with high 
probability. This is higher than the number of iterations of state of the art 
interior point methods, but lower than the overall complexity
$\cO(n^{3.5} \log n)$ of the high probability estimate mentioned above, which 
includes the complexity of an individual operation.

In this paper we consider a flow for linear programming  
proposed by Faybusovich \cite{faybusovich}, for which $F(x)$ is given by 
(\ref{field}). Substituting (\ref{field}) into the general equation 
(\ref{contsys}) we obtain (\ref{dynamics}), which realizes the 
Faybusovich algorithm for LP. We consider real inputs that are
drawn from a Gaussian probability distribution. For any feasible instance of 
the LP problem, the flow converges to the solution. We consider the 
question: Given the probability distribution of LP instances, 
what is the probability distribution of the convergence rates to the 
solution? The convergence rate measures the asymptotic computation time:
the time to reach an $\epsilon$ vicinity of the attractor, where
$\epsilon$ is arbitrarily small. The main result of this paper, as stated in 
Theorem (\ref{mainresult}), is that with high probability and on the average,
the asymptotic computation time is ${\cal O}\left(\sqrt{n}|\log 
\epsilon|\right)$,  where $n$ is the problem size and $\epsilon$ is the 
required precision (see also Corollary (\ref{corollary-deltaHighProb})).

In practice, the solution to arbitrary precision is not always required,
and one may need to know only whether the flow (\ref{contsys}) or 
(\ref{dynamics}) has reached the vicinity of the optimal vertex, 
or which vertex out of a given set of vertices will be approached by the 
system. Thus, the non-asymptotic behavior of the flow needs to be considered
\cite{dds2}. In this case, only a heuristic estimate of the 
computation time is presented, and in Section 6 we conjecture that the 
associated complexity is ${\cal O}\left(n\log n\right)$, as mentioned above.

The rest of the paper is organized as follows: In section 2 the 
Faybusovich flow is presented along with an expression for its convergence 
rate. The probabilistic ensemble of the LP instances is presented in 
section 3. The distribution of the convergence rate of this flow
is calculated analytically in the framework of random matrix theory (RMT) 
in section 4. In secton 5 we introduce the concept of ``high-probability 
behavior'' and use the results of section 4 to quantify the high-probability 
behavior of our probabilistic model.  In section 6 we provide measures of 
complexity when precision asymptotic in 
$\epsilon$ is not required. Some of the results in sections 6.2-8 are
heuristic, supported by numerical evidence. The structure of the 
distribution functions of parameters that 
control the convergence is described in section 7 and its numerical 
verification is presented in section 8.  Finally, the results of this work 
and their possible implications are discussed in section 9.
Some technical details are relegated to the appendices. Appendix A 
contains more details of the Faybusovich flow. Appendix B exposes the 
details of the analytical calculation of the results presented in section 4, 
and appendix C contains the necessary details of random matrix theory 
relevant for that calculation.

\section{A flow for linear programming}

We begin with the definition of the linear programming problem (LP)
and a vector field for solving it introduced 
by Faybusovich in \cite{faybusovich}.
The {\it standard form} of LP is to find
\begin{equation}
\label{standard}
\max \{ c^{T}x~:~ x\in \R ^{n}, 
  A x =  b,x \geq 0  \}
\end{equation}
where $c \in \R^n, b \in \R^m, 
A \in \R^{m \times n}$ 
and $m\leq n$.  
The set generated by the constraints in (\ref{standard}) is a polyheder.
If a bounded optimal solution exists, it is obtained at one of its vertices.
Let ${\cal B} \subset \{1,\ldots,n\}$, $|{\cal B}|=m$, and
${\cal N} =  \{1,\ldots,n\} \setminus {\cal B}$, and denote by $x_{{\cal B}}$ the coordinates with
indices from ${\cal B}$, and by 
$A_{\cal B}$, the $m \times m$ matrix whose columns are
the columns of $A$ with indices from ${\cal B}$.
A vertex of the LP problem is defined by a set of indices ${\cal B}$, 
which is called a {\em basic set}, if 
\begin{equation}\label{xB}
 x_{\cal B} = A_{\cal B}^{-1}  b \geq  0 \;.
\end{equation}
The components of a vertex are $x_{\cal B}$ that satisfy (\ref{xB}), and $x_{\cal N} = 0$. 
The set ${\cal N}$ is then called a {\em non-basic} set.
Given a vector field that converges to an optimal solution represented
by basic and non-basic sets ${\cal B}$ and ${\cal N}$, its solution
$x(t)$ can be decomposed as $(x_{\cal N}(t), x_{\cal B}(t))$ where
$x_{\cal N}(t)$ converges to 0, and $x_{\cal B}(t)$ converges to
$A_{\cal B}^{-1} b$. 

In the following we consider the non-basic set 
${\cal N}=\{1,\ldots,n-m\}$, and for notational convenience
denote the $m\times m$ matrix $A_{\cal B}$ by $B$ and denote
$A_{\cal N}$ by $N$, i.e. $A=(N,B)$.

The Faybusovich vector field  is a projection of the gradient of the linear 
cost function onto the constraint set, relative to a Riemannian metric which
enforces the positivity constraints $x\geq 0$ \cite{faybusovich}.
Let $h(x) = c^T x$. We denote this projection by $\mbox{grad}\, h$.
The explicit form of the gradient is:
\begin{equation} \label{field}
	\mbox{grad}\, h(x)=[X - X A^{T}
		(A X A^{T})^{-1} A X]\: c\; ,
\end{equation}
where $X$ is the diagonal matrix $\mbox{Diag}(x_1 \dots x_n)$.
It is clear from (\ref{field}) that 
$$A\mbox{grad}\, h(x) =0\,.$$ 
Thus, the dynamics 
\beq\label{dynamics}
{dx\over dt} = \mbox{grad}\, h(x)
\eeq
preserves the constraint $Ax=b$ in (\ref{standard}).
Thus, the faces of the polyheder are invariant
sets of the dynamics induced by $\mbox{grad}\, h$. Furthermore, it is 
shown in \cite{faybusovich} that the fixed 
points of $\mbox{grad}\,h$ coincide with the vertices of the polyheder, 
and that the dynamics converges exponentially to 
the maximal vertex of the LP problem.
Since the formal solution of the Faybusovich vector field is the basis
of our analysis we give its derivation in Appendix A.

Solving (\ref{dynamics}) requires an appropriate initial condition - an 
interior point in this case. This can be addressed either by using the 
``big-M'' method \cite{Saigal}, which has essentially the same convergence 
rate, or by solving an auxiliary linear programming problem 
\cite{Ye-highprob}. We stress that here, the initial interior point 
is not an input for the computation, but rather a part of the algorithm. 
In the analog implementation the initial point should be found by the same 
device used to solve the LP problem.

The linear programming problem (\ref{standard}) has $n-m$ independent 
variables. The formal solution shown below, describes the time evolution 
of the $n-m$ variables $x_{\cal N}(t)$, in terms of the variables 
$x_{\cal B}(t)$. When ${\cal N}$ is the non-basic set of an optimal 
vertex of the LP problem, $x_{\cal N}(t)$ converges to 0, and 
$x_{\cal B}(t)$ converges to $A_{\cal B}^{-1}b$.
Denote by $e^1,\ldots,e^n$ the standard basis of $\R^n$, and define
the $n-m$ vectors 
\begin{equation}\label{mu}
\mu^i = e^i + \sum_{j=1}^{m} \alpha_{ji} e^j~,
\end{equation}
where 
\begin{equation}\label{alpha}
\alpha_{ji}= - (B^{-1} N)_{ji} 
\end{equation}
is an $m \times (n-m)$ matrix.
The vectors $\mu^i$ are perpendicular to the 
rows of $A$ and are parallel to the faces of the 
polyheder defined by the constraints.
In this notation the analytical solution is (see Appendix \ref{app-solution}):
\begin{equation}
\label{solution}
x_i(t) = x_i(0) \exp \left( -\Delta_i t -
\sum_{j=1}^{m} \alpha_{ji} \log \frac{x_{j+n-m}(t)}{x_{j+n-m}(0)} \right) ~,~~i\in \cn =
\{1,\ldots,n-m\}
\end{equation}
where $x_i(0)$ and $x_{j+n-m}(0)$ are components of the initial 
condition, $x_{j+n-m}(t)$ are the $x_{\cb}$ components of the solution, 
and
\begin{equation}\label{deltas}
\Delta_i = - < \mu^i , c>  = -c_i  - \sum_{j=1}^{m} c_j \alpha_{ji}   
\end{equation}
(where $<.,.>$ is the Euclidean inner product).

An important property which relates the signs of the $\Delta_i$ and 
the optimality of the partition of $A$ (into $(B,N)$) relative to 
which they were computed is now stated:
\begin{lemma}\cite{faybusovich}
For a polyhedron with $\{n-m+1,\ldots,n\}$, a basic set of
a maximum vertex,
$$\Delta_i \geq 0~~i=1,\ldots,n-m\;.$$
\end{lemma}
The converse statement does not necessarily hold. The $\Delta_i$ are 
independent of $b$. Thus we may have that all $\Delta_i$ are positive, 
and yet the constraint set is empty.
\begin{remark}

Note that the analytical solution is only a formal one, and does not 
provide an answer
to the LP instance, since the $\Delta_i$ depend on the partition of $A$, and
only relative to a partition corresponding to a 
maximum vertex are all the $\Delta_i$ positive.
\end{remark}

The quantities $\Delta_i$ are the convergence rates of the Faybusovich flow,
and thus measure the time required to reach the $\epsilon$-vicinity of the 
optimal vertex, where $\epsilon$ is arbitrarily small:
\beq\label{tepsilon}
T_{\epsilon} \sim {|\log \epsilon | \over \Delta_{min}}\,,
\eeq
where 
\begin{equation}\label{Deltamin}
\Delta_{\min} = \min_i \Delta_i\;.
\end{equation}
Therefore, if the optimal vertex is required with arbitrary precision 
$\epsilon$ , then the computation time (or complexity) is ${\cal O}
\left(\Delta_{min}^{-1}|\log \epsilon |\right)$.

In summary, if the $\Delta_i$ are small then large computation times will 
be required.
The $\Delta_i$ can be arbitrarily small when the inputs are real numbers, 
resulting in an unbounded computation time.
However, we will show that in the probabilistic model, 
which we define in the next section, ``bad'' instances are rare, and the 
flow performs well ``with high probability'' (see Theorem (\ref{mainresult})
and Corollary (\ref{corollary-deltaHighProb})).

\section{The probabilistic model}

We now define the ensemble of LP problems for which we analyze the
complexity of the Faybusovich flow.
Denote by $N(0,\si^2)$ the standard Gaussian distribution with 0 mean and 
variance $\si^2$.
Consider an ensemble in which the components of $(A,b,c)$ are i.i.d.
(independent identically distributed) random variables
with the distribution $N(0,\si^2)$.
The model will consist of the following set of problems:
\begin{eqnarray}\label{LPM}
LPM ~ = ~ \{ (A,b,c)  & | & (A,b,c) \mbox{~are i.i.d. variables with the distribution~} 
N(0,\si^2)~~~~~~~ \\ \nonumber
 & &\mbox{and the LP problem has a bounded optimal solution} \} \,.
\end{eqnarray}
Therefore, we use matrices with 
a distribution $N(0,\sigma^2)$:
\begin{equation}\label{aens}
f(A) = {1\over {\cal Z}_A} \exp \left( -\var\rmtr A^TA \right)
\end{equation}
with normalization 
\begin{equation}\label{anorm}
{\cal Z}_A = \int d^{mn}A ~\exp \left( -\var\rmtr A^TA \right) = 
\ivar^{mn/2}\,. 
\end{equation}

The ensemble (\ref{aens}) factorizes into $mn$ 
i.i.d. Gaussian random variables for each of the 
components of $A$.

The distributions of the vectors $c$ and $b$ are defined by:
\begin{equation}\label{cens}
f(c) = {1\over {\cal Z}_c} \exp \left( -\var c^Tc \right)
\end{equation}
with normalization 
\begin{equation}\label{cnorm}
{\cal Z}_c = \int d^nc~ \exp \left( -\var c^Tc \right)
= \ivar^{n/2} \,,
\end{equation}
and
\begin{equation}\label{bens}
f(b) = {1\over {\cal Z}_b} \exp \left( -\var b^Tb \right)
\end{equation}
with normalization 
\begin{equation}\label{bnorm}
{\cal Z}_b = \int d^mb~ \exp \left( -\var b^Tb \right)
= \ivar^{m/2}\,.
\end{equation}

With the introduction of a probabilistic model of LP instances 
$\Delta_{\min}$ becomes a random variable.
We wish to compute the probability distribution of $\Delta_{\min}$
for instances with a bounded solution, when $\Delta_{\min} > 0$.
We reduce this problem to the simpler task of
computing ${\cal P}(\Delta_{min} > \Delta | \Delta_{min} 
> 0)$, in which the condition $\Delta_{min} > 0$ is much easier to 
impose than the condition that an instance produces an LP problem with a 
bounded solution. This reduction is justified by the following lemma:
\begin{lemma}\label{lemma-noncorrelation}
\begin{eqnarray}\label{indcon}
& {\cal P}(\Delta_{\min} > \Delta |\mbox{LP instance has a bounded maximum 
vertex}) = \\ \nonumber
& {\cal P}(\Delta_{\min} > \Delta | \Delta_{\min} > 0).
\end{eqnarray}
\end{lemma}

\begin{proof}
Let $(A,b,c)$ be an LP instance chosen according to the probability 
distributions (\ref{aens}), (\ref{cens}) and (\ref{bens}).
There is a {\em unique} orthant (out of the $2^n$ orthants) 
where the constraint set $Ax=b$ defines a nonempty polyheder.
This orthant is not necessarily the positive orthant, as in the standard
formulation of LP.

Let us consider now {\em any} vertex of this polyheder with
basic and non-basic sets ${\cal B}$ and ${\cal N}$. 
Its $m$ non-vanishing coordinates $x_{\cal B}$ are given by solving 
$A_{\cal B} x_{\cal B} = b $. The matrix $A_{\cal B}$ is full rank with 
probability 1; also, the components of $x_{\cal B}$ are non-zero and finite 
with probability 1. Therefore, in the probabilistic analysis we can assume 
that $x_{\cal B}$ is well defined and non-zero. With this vertex we 
associate the $n-m$ quantities  $\Delta_i = -(c_{\cal N})_i  + 
(c_{\cal B}^T A_{\cal B}^{-1}A_{\cal N})_i$, from (\ref{deltas}).    

We now show that there is a set of $2^m$ {\it equiprobable} 
instances, which contains the instance $(A,b,c)$, that shares the 
same vector $b$ and the same values of $\{\Delta_i\}$, when computed 
according to the given partition. This set contains a unique instance with 
$x_{\cal B}$ in the positive orthant. Thus, if $\Delta_{min}>0$, the latter 
instance will be the unique member of the set which has a bounded optimal 
solution.

To this end, consider the set ${\cal R}(x_{\cal B})$ of the $2^m$ 
reflections ${\cal Q}_l x_{\cal B}$ of $x_{\cal B}$, where ${\cal Q}_l$ 
is an $m\times m$ diagonal matrix with diagonal entries $\pm 1$ and 
$l=1, 2, .... ,2^m$.

Given the instance $(A,b,c)$ and a particular partition into basic and 
non-basic sets, we split $A$ columnwise into $(A_{\cal B},A_{\cal N})$ and 
$c$ into $(c_{\cal B},c_{\cal N})$. 
Let ${\cal S}$ be the set of $2^m$  instances
$((A_{\cal B}{\cal Q}_l,A_{\cal N}), b, ({\cal Q}_lc_{\cal B},c_{\cal 
N}))$ where $l=1,\ldots,2^m$. The vertices ${\cal Q}_l x_{\cal B}$ of these 
instances, which correspond to the prescribed partition, comprise the 
set ${\cal R}(x_{\cal B})$, since $(A_{\cal B}{\cal Q}_l)({\cal Q}_l 
x_{\cal B}) = b $. Furthermore, all elements in ${\cal R}(x_{\cal B})$ 
(each of which corresponds to a different instance) have the same set of 
$\Delta$'s, since $\Delta_i = -(c_{\cal N})_i  + [({\cal Q}_lc_{\cal 
B})^T (A_{\cal B}{\cal Q}_l)^{-1}A_{\cal N}]_i$. 
Because of the symmetry of the  ensemble under the reflections ${\cal 
Q}_l$, the probability of all instances in ${\cal S}$ is the same.

All the vertices belonging to ${\cal R}(x_{\cal B})$ have the same 
$\Delta_i$'s with the same probability, and exactly one is in the positive 
orthant. Thus, if $\Delta_{\min} > 0$, the latter vertex 
is the unique element from ${\cal S}$ which is the optimal vertex 
of an LP problem with a bounded solution. Consequently, the probability of 
having any prescribed set of $\Delta_i$'s, and in particular, the probability 
distribution for the $\Delta_i$'s given $\Delta_{min} > 0$, is not 
affected by the event that the LP instance has a bounded optimal solution 
(i.e., that the vertex is in the positive orthant). In other words, these 
are independent events. Integration over all instances and taking this way 
into account all possible sets ${\cal S}$ while imposing the requirement
$\{\Delta_{\min} > \Delta | \Delta_{\min} > 0\}$ results in 
(\ref{indcon}).
\end{proof}

The event $\Delta_{\min} > 0$ corresponds to a specific partition of $A$ into
basic and non-basic sets ${\cal B,N}$, respectively.
It turns out that it is much easier to analytically calculate the probability 
distribution of $\Delta_{min}$ for a given partition of the matrix $A$.
It will be shown in what follows that in the probabilistic model we defined,
${\cal P}(\Delta_{\min} > \Delta | \Delta_{\min} > 0)$ is proportional
to the probability that $\Delta_{\min} > \Delta$ for a fixed partition.
Let $W_j$ be the event that a partition $j$ of the matrix $A$
is an optimal partition, i.e. all $\Delta_i$ are
positive (j is an index with range $1,\ldots,$ $\!\!\! n\choose m$).
Let the index 1 stand for the partition where $B$ is taken from the last
$m$ columns of $A$.
We now show:
\begin{lemma}\label{lemma-partition}
Let $\Delta > 0$ then
$$
{\cal P}(\Delta_{min}> \Delta | \Delta_{min} > 0) = 
{\cal P}(\Delta_{min}> \Delta | W_1) \;.
$$
\end{lemma}
 
\begin{proof}
Given that $\Delta_{\min} > 0$, there is a unique optimal
partition since a non-unique optimal partition occurs only if $c$ is orthogonal
to some face of the polyhedron, in which case $\Delta_i=0$ for some $i$.
Thus we can write:
\begin{eqnarray}
{\cal P}(\Delta_{min} > \Delta | \Delta_{min} > 0) & = & 
\sum_{j} {\cal P}(\Delta_{min} > \Delta | \Delta_{min} > 0, W_j) {\cal P}(W_j)  \\
& = & \sum_{j} {\cal P}(\Delta_{min} > \Delta | W_j) {\cal P}(W_j)  \;,
\end{eqnarray}
where the second equality holds since the event $W_j$ is contained in the
event that $\Delta_{min} > 0$.
The probability distribution of $(A,c)$ is invariant under permutations of columns
of $A$ and $c$, and under permutations of rows of $A$.
Therefore the probabilities 
${\cal P}(W_j)$ are all equal, and so are
${\cal P}(\Delta_{min}> \Delta | \Delta_{min}>0, W_j)$, and the result follows.

\end{proof}

We define
\begin{equation}
\Delta_{min1} = \min \{\Delta_i ~|~ \Delta_i 
\mbox{~are computed relative to the partition 1} \}
\end{equation}
Note that the definition of $\Delta_{\min}$ in equation (\ref{Deltamin})
is relative to the optimal partition.
To show that all computations can be carried out
for a fixed partition of $A$ we need the next lemma:

\begin{lemma}\label{lemma-symmetry}
Let $\Delta > 0$ then
$$
{\cal P}(\Delta_{\min} > \Delta | \Delta_{\min} > 0) =
\frac{{\cal P}(\Delta_{min1} > \Delta)}{{\cal P}(\Delta_{min1} > 0)} \;.
$$
\end{lemma}

\begin{proof}
The result follows from
\begin{equation}
{\cal P}(\Delta_{\min} > \Delta | W_1) = 
{\cal P}(\Delta_{min1} > \Delta | \Delta_{min1} > 0),
\end{equation}
combined with the result of the previous lemma and the definition of
conditional probability.
\end{proof}

In view of the symmetry of the joint probability distribution (j.p.d.)
of $\Delta_1 ,\ldots, \Delta_{n-m}$, given by (\ref{jpd1}) and 
(\ref{jpd11}), the normalization constant ${\cal P}(\Delta_{min1}>0)$ 
satisfies:
\begin{equation}\label{Ppositive}
{\cal P}(\Delta_{min1}>0) = 1/2^{n-m} \;.
\end{equation}

\begin{remark}\label{remark-degeneracy}
Note that we are assuming throughout this work, that the optimal vertex is 
unique, i.e.,
given a partition $(\cn,\cb)$ of $A$ that corresponds to an optimal vertex,
the basic components are all non-zero. The reason is that if one of the 
components of the optimal vertex vanishes, all of its permutations with the 
$n-m$ components of the non-basic set result 
in the same value of $c^Tx$. Vanishing of one of the components of the 
optimal vertex requires that $b$ is a linear combination of columns of $A$, 
that is an event of zero measure in our probabilistic ensemble. Therefore 
this case will not be considered in the present work.
\end{remark}

\section{Computing the distributions of $\Delta_{min1}$ and of $\Delta_{min}$}

In the following we compute first the distribution of $\Delta_{min1}$
and use it to obtain the distribution of $\Delta_{min}$ via Lemma 
(\ref{lemma-symmetry}).
We denote the first $n-m$ components of $c$ by $y$, and its last
$m$ components by $z$.
In this notation equation (\ref{deltas}) for 
$\Delta_i$ takes the form:
\begin{equation}\label{delta}
\Delta_p = - y_p + (z^T B^{-1} N)_p ~~p=1,\ldots,n-m \,.
\end{equation}
Our notation will be such that indices 
$$i,j,k,\ldots\quad\quad {\rm range~over}\quad\quad  
1,2,\ldots,m$$ 
and
$$p,q,\ldots\quad\quad {\rm range~over}\quad\quad 1,2,\ldots, n-m.$$ 
In this notation, the ensembles (\ref{aens}) and (\ref{cens}) may be written as
\begin{eqnarray}\label{ensembles}
f(A) &=& f(N,B) = {1\over {\cal Z}_A} \exp \left[ 
-\var \left(\sum_{ij} B_{ij}^2 + \sum_{ip} N_{ip}^2\right) \right]
\nonumber\\
\\
f(c) &=& f(y,z) = {1\over {\cal Z}_c} \exp \left[
-\var \left(\sum_i z_i^2 + \sum_p y_p^2\right) \right] \,\nonumber.
\end{eqnarray}

We first compute 
the joint probability distribution (j.p.d.) of $\Delta_1,\ldots,\Delta_{n-m}$
relative to the partition 1.
This is denoted by $f_1(\Delta_1,\ldots,\Delta_{n-m})$.
Using (\ref{delta}), we write 
\begin{eqnarray}\label{jpddelta}
f_1(\Delta_1,\ldots,\Delta_{n-m}) &=& \int  d^{m^2}B
 \, d^{m(n-m)}N\,d^mz\,d^{n-m}y\nonumber\\
&& f(N,B)f(y,z) \prod_{q=1}^{n-m}
\delta \left(\Delta_q + y_q - \sum_{i,j=1}^m z_j(B^{-1})_{ji}N_{iq}\right)\,,
\end{eqnarray}
where $\delta(x)$ is the Dirac delta function.
We note that this j.p.d. is 
not only completely symmetric under permuting the $\Delta_p$'s, but 
is also 
{\em independent} of the partition relative to which it is computed.

We would like now to perform the integrals in (\ref{jpddelta}) and 
obtain a more explicit expression for $f_1(\Delta_1,\ldots,\Delta_{n-m})$. 
It turns out that direct integration over the $y_q$'s, using the $\delta$ 
function, is not the most efficient way to proceed. 
Instead, we represent each of the $\delta$ functions as a Fourier integral. 
Thus,
\begin{eqnarray*}
f_1(\Delta_1,\ldots,\Delta_{n-m}) &=& \int
d^{m^2}B \, d^{m(n-m)}N \,d^mz\,d^{n-m}y 
~{d^{n-m}\lambda\over (2\pi)^{n-m}}~f(N,B)f(y,z) \nonumber\\
&&\cdot \exp \left[ ~i\sum_q \lambda_q \left(\Delta_q + y_q - 
\sum_{i,j=1}^m z_j(B^{-1})_{ji}N_{iq}\right) \right] \,.
\end{eqnarray*}

Integration over $N_{ip}, \lambda_q$ and $y_p$ is straight forward and yields 
\begin{eqnarray}\label{jpd1}
f_1(\Delta_1,\ldots,\Delta_{n-m}) = &
\iivar^{{m^2+n\over 2}}\int {{d^{m^2}B \, 
d^mz} \over {\left[z^T(B^TB)^{-1}z + 1\right]^{{n-m\over 2}}}}
\nonumber\\
&\cdot \exp \left[ ~-\var \left(\sum_{ij} B_{ij}^2 + \sum_i z_i^2
+ {\sum_p\Delta_p^2\over z^T(B^TB)^{-1}z + 1}\right) \right] \,.
\end{eqnarray}
Here the complete symmetry of $f_1(\Delta_1,\ldots, \Delta_{n-m})$ under
permutations of the $\Delta_p$'s is 
explicit, since it is a function of $\sum_p\Delta_p^2$.

The integrand in (\ref{jpd1}) contains the combination 
\begin{equation}\label{uvariable}
u(B,z) = {1\over z^T(B^TB)^{-1}z + 1}\,.
\end{equation}
Obviously, $0\leq u(B,z)\leq 1$. It will turn out to be very useful to consider
the distribution function $P(u)$ of the random variable $u = u(B,z)$, namely,  
\begin{equation}\label{udistribution}
P(u) = 
\iivar^{{m^2+m\over 2}}\int d^{m^2}B \, 
d^mz\, e^{-\var \left(\rmtr\, B^TB + z^Tz \right)}
\cdot \delta \left(u - {1\over z^T(B^TB)^{-1}z + 1}\right)\,.
\end{equation}

Note from (\ref{uvariable}) that $u(\lambda B, \lambda z) = u(B,z)$. Thus, 
in fact, $P(u)$ is independent of the (common) variance $\si$ of the Gaussian 
variables $B$ and $z$, and we might as well rewrite (\ref{udistribution}) 
as 
\begin{equation}\label{udistribution1}
P(u) = 
\left({\lambda\over\pi}\right)^{{m^2+m\over 2}}\int d^{m^2}B \, 
d^mz\, e^{-\lambda\left(\rmtr\, B^TB + z^Tz \right)}
\cdot \delta \left(u - {1\over z^T(B^TB)^{-1}z + 1}\right)\,,
\end{equation}
with $\lambda>0$ an arbitrary parameter.

Thus, if we could calculate $P(u)$ explicitly, we would be able to 
express the j.p.d. $f_1(\Delta_1,\ldots,\Delta_{n-m})$ in (\ref{jpd1}) 
in terms 
of the one dimensional integral
\begin{equation}\label{jpd11}
f_1(\Delta_1,\ldots,\Delta_{n-m}) = 
\iivar^{{n-m\over 2}}\int\limits_0^\infty\,du\, P(u)\, 
u^{{n-m\over 2}}\,
\exp \left[-\var \left(u\sum_{p=1}^{n-m}\Delta_p^2\right)\right]\,,
\end{equation}
as can be seen by comparing (\ref{jpd1}) and (\ref{udistribution}).

In this paper we are interested mainly in the minimal $\Delta$. Thus, we 
need $f_{min1}(\Delta)$, the probability density 
of $\Delta_{\min1}$. Due to the symmetry of 
$f_1(\Delta_1, \ldots, \Delta_{n-m})$, which is explicit 
in (\ref{jpd11}), we can express $f_{min1}(\Delta)$ simply as 
\begin{equation}\label{pmin}
f_{min1} (\Delta) = (n-m)\int\limits_\Delta^\infty\,d\Delta_2\,
\ldots d\Delta_{n-m} 
f_1 (\Delta,\Delta_2,\ldots, \Delta_{n-m})\,.
\end{equation}

It will be more convenient to consider the complementary cumulative 
distribution (c.c.d.)
\begin{equation}\label{fQrelation}
{\cal Q}(\Delta) = {\cal P}(\Delta_{min1} > \Delta) 
= \int\limits_\Delta^\infty f_{min1}(u) du\,,
\end{equation}
in terms of which 
\begin{equation}\label{pmin1}
f_{min1}(\Delta) = - {\partial \over\partial \Delta} {\cal Q} (\Delta) ~.
\end{equation}
The c.c.d. ${\cal Q}(\Delta)$ may be expressed as a symmetric integral 
\begin{eqnarray}\label{fQrelationsym}
{\cal Q}(\Delta) 
=\int\limits_\Delta^\infty\,d\Delta_1\,\ldots d\Delta_{n-m} 
f_1 (\Delta_1,\Delta_2,\ldots, \Delta_{n-m})
\end{eqnarray}
over the $\Delta$'s, and thus, it is computationally a more convenient object
to consider than $f_{min1}(\Delta) $. 

From (\ref{fQrelationsym}) and (\ref{jpd11}) we obtain that 
\begin{equation}\label{cumulant1}
{\cal Q}(\Delta) = \iivar^{{n-m\over 2}}
\int\limits_0^\infty\,du\, P(u)\, 
\left(\sqrt{u}\,\int\limits_\Delta^\infty\,dv\, e^{-\var\,u v^2}
\right)^{n-m}\,,
\end{equation}
and from (\ref{cumulant1}) one readily finds that
\begin{equation}\label{cumulant0}
{\cal Q}(0) = {1\over 2^{n-m}}\,,
\end{equation}
(as well as ${\cal Q} (-\infty) = 1$, by definition of ${\cal Q}$).

Then, use of the integral representation 
\begin{equation}\label{error}
1- {\rm erf}(x) = {\rm erfc}(x) = {2\over\sqrt{\pi}}
\int\limits_x^\infty dv~e^{-v^2}\,,
\quad\quad(x>0)
\end{equation}
and (\ref{cumulant0}) leads (for $\Delta>0$) to
\begin{eqnarray}\label{Qfinal}
{\cal Q}(\Delta) = 
{\cal Q}(0)\,
\int\limits_0^\infty\,du\, P(u)\, 
\left({\rm erfc}\,[\Delta\,\sqrt{{u\over 2\si^2}}\,]\right)^{n-m}\,.
\end{eqnarray}
This expression is an {\em exact} integral representation of ${\cal Q} (\Delta)$
(in terms of the yet undetermined probability distribution $P(u)$).

In order to proceed, we have to determine $P(u)$. Determining
$P(u)$ for any pair of integers $(n,m)$ in (\ref{udistribution1}) in a closed 
form is a difficult task. However, since we are interested mainly in the 
asymptotic behavior of computation times,
we will contend ourselves in analyzing the behavior of 
$P(u)$ as $n,m\rightarrow\infty$, with 
\beq\label{r}
r\equiv m/n <1
\eeq
held fixed.

We were able to determine the large $n,m$ behavior of 
$P(u)$ (and thus of $f_1 (\Delta_1,\Delta_2,\ldots, \Delta_{n-m})$ and 
${\cal Q}(\Delta)$) using standard methods \cite{bipz, rmt} of random matrix 
theory \cite{mehta}.

This calculation is presented in detail in Appendix B. We show there
(see Eq. (\ref{puasymptotic})) that the leading asymptotic behavior of 
$P(u)$ is   
\begin{equation}\label{pexplicit}
P(u) = \sqrt{{m\over 2\pi u}}\, e^{-{mu\over 2}}\,,
\end{equation}  
namely, $\sqrt{u}$ is simply a Gaussian variable, with variance proportional
to $1/\sqrt{m}$. Note that (\ref{pexplicit}) is independent of the 
width $\si$, which is consistent with the remark preceding 
(\ref{udistribution1}).

Substituting (\ref{pexplicit}) in (\ref{jpd11}), 
we obtain, with the help of the integral representation 
\beq\label{eulergamma}
\Gamma (z) = \int\limits_0^\infty t^{z-1}\, e^{-t} \,dt
\eeq
of the $\Gamma$ function, the large $n,m$ behavior of the j.p.d.
$f_1(\Delta_1,\ldots,\Delta_{n-m})$ as 
\begin{equation}\label{jpdasymptotic}
f_1(\Delta_1,\ldots,\Delta_{n-m}) = 
\sqrt{m}\si \,\Gamma\left({n-m+1\over 2}\right)\,
\left({1\over\pi}\,{1\over m\si^2 + \sum_p\Delta_p^2
}\right)^{{n-m+1\over 2}}\,.
\end{equation}
Thus, the $\Delta$'s follow asymptoticly a multi-dimensional Cauchy
distribution. It can be checked that (\ref{jpdasymptotic}) is properly 
normalized to 1.

Similarly, by substituting (\ref{pexplicit}) in (\ref{Qfinal}), and
changing the variable to $y=\sqrt{mu/2}$, we obtain 
the large $n,m$ behavior of 
${\cal Q}(\Delta)$ as 
\begin{equation}\label{Qasymptotic}
{\cal Q}(\Delta) = 
{2{\cal Q}(0)\over\sqrt{\pi}}\,
\int\limits_0^\infty\,dy\,e^{-y^2}  
\left({\rm erfc}\,\left[\Delta\,{y\over \sqrt{m}\si}\,\right]
\right)^{n-m}\,.
\end{equation}
As a consistency check of our large $n,m$ asymptotic 
expressions, we have verified, with the help of (\ref{eulergamma}), that substituting 
(\ref{jpdasymptotic}) into (\ref{fQrelationsym}) leads to (\ref{Qfinal}), 
with $P(u)$ there given by the asymptotic expression (\ref{pexplicit}).

We are interested in the scaling behavior of ${\cal Q}(\Delta)$ in 
(\ref{Qasymptotic}) in the limit $n,m\rightarrow\infty$. In this large 
$n,m$ limit, the factor  
$\left({\rm erfc}\,\left[\Delta\,{y\over \sqrt{m}\si}\,\right]
\right)^{n-m}$ in (\ref{Qasymptotic}) decays rapidly to zero. Thus, 
the integral in (\ref{Qasymptotic}) will be appreciably different from zero 
only in a small region around $\Delta=0$, where the erfc function is very 
close to 1. More precisely, using ${\rm erfc}\, x = 1 - {2x\over \sqrt{\pi}} + 
{\cal O}(x^2)$, we may expand the erfc term in (\ref{Qasymptotic}) as 
\begin{equation}\label{erfcterm}
\left({\rm erfc}\,\left[\Delta\,{y\over \sqrt{m}\si}\,\right]\right)^{n-m}= 
\left(1 - {2y\Delta\over \sqrt{\pi m\si^2}} + \cdots
\right)^{n-m}
\end{equation}
(due to the Gaussian damping factor in (\ref{Qasymptotic}), this expansion 
is uniform in $y$). Thus, we see that ${\cal Q}(\Delta)/{\cal Q}(0)$ will be 
appreciably different from zero only for values of $\Delta/\si$ of 
the order up to $1/\sqrt{m}$, for which (\ref{erfcterm}) exponentiates into a 
quantity of 
${\cal O}(1)$, and thus
\begin{equation}\label{Qasymptoticscaling}
{\cal Q}(\Delta) \simeq 
{2{\cal Q}(0)\over\sqrt{\pi}}\,
\int\limits_0^\infty\,dy\,e^{-y^2}  
\exp\left(-{2\over\sqrt{\pi}}\left({n\over m}-1\right)\,y\delta\right)\,,
\end{equation}
where
\begin{equation}
\delta={\sqrt{m}\Delta\over\si}
\end{equation}
is ${\cal O} (m^0)$. Note that $m/n$ is kept finite and fixed. 
The integral in 
(\ref{Qasymptoticscaling}) 
can be done, and thus we arrive at the explicit scaling behavior of
the c.c.d.
\begin{equation}\label{scalingQ}
{\cal Q}(\Delta) =
{\cal Q}(0)\, e^{x_\Delta^2}\,{\rm erfc}(x_\Delta)\,,
\end{equation}
where
\begin{equation}\label{scalingvariable}
x_\Delta = \eta_\Delta (n,m) \Delta\,,
\end{equation}
with
\begin{equation}\label{etadelta}
\eta_\Delta (n,m) =  {1\over\sqrt{\pi}}\left({n\over m}-1\right)\,{\sqrt{m}\over\sigma}\,.
\end{equation}
The c.c.d. ${\cal Q}(\Delta)$ depends, in principle, on all the three 
variables $n,m$ and $\Delta$. The result (\ref{scalingQ}) demonstrates, that 
in the limit $(n,m)\rightarrow\infty$ (with $r=m/n$ held finite and fixed), 
${\cal Q}(\Delta)$ is a function only 
of {\em one scaling variable}: the $x_\Delta$ defined in (\ref{scalingvariable}).

We have compared (\ref{scalingQ}) and (\ref{scalingvariable}) against results
of numerical simulations, for various values of $n/m$. The results are 
shown in Figures 2 and 3 in Section 8.

Establishing the explicit scaling expression of the probability 
distribution of the convergence rate constitutes the main result in our 
paper, which we summarize by the following Theorem:

\begin{theorem}\label{mainresult}
Assume that LP problems of the form (\ref{standard}), with the instances 
distributed according to (\ref{aens})-(\ref{bnorm}), are solved by the 
Faybusovich algorithm (\ref{dynamics}). Then, in the asymptotic limit 
$n\rightarrow\infty$, $m\rightarrow\infty$ with $0< r=m/n < 1$ kept fixed, 
the convergence rate $\Delta_{min}$ defined by (\ref{Deltamin}) is 
distributed according to 
\begin{equation}\label{mainresulteq}
{\cal P}(\Delta_{min} > \Delta| \mbox{bounded optimal solution} )= 
e^{x_\Delta^2}\,{\rm erfc}(x_\Delta)\,,
\end{equation} 
where $x_{\Delta}$ is given by (\ref{scalingvariable}). 
\end{theorem}
\begin{proof}
${\cal Q}(\Delta) = {\cal P}(\Delta_{min1} > \Delta)$ by (\ref{fQrelation}).  
Therefore, use of (\ref{Ppositive}) and (\ref{cumulant0}), namely, 
$${\cal P}(\Delta_{min1}>0)={\cal Q}(0) = 1/2^{n-m}$$, and of (\ref{scalingQ}) 
implies 
\beq\label{Pdeltamin11}
{\cal P}(\Delta_{min1}>\Delta ) = {1\over 2^{n-m}}\, 
e^{x_\Delta^2}\,{\rm erfc}(x_\Delta)\,, 
\end{equation}
but according to Lemma (\ref{lemma-symmetry}), 
$$
{\cal P}(\Delta_{\min} > \Delta | \Delta_{\min} > 0) =
\frac{{\cal P}(\Delta_{min1} > \Delta)}{{\cal P}(\Delta_{min1} > 0)} \;.
$$
Finally, substituting (\ref{Pdeltamin11}) and (\ref{Ppositive}) in the last 
equation, and use of Lemma (\ref{lemma-noncorrelation}), leads to the 
statement of the theorem.
\end{proof}

From (\ref{scalingQ}) and (\ref{scalingvariable}), we can obtain 
the probability density $f_{min1}(\Delta)$ of $\Delta_{\min1}$, using
(\ref{pmin1}). In particular, we find 
\begin{equation}\label{zerodensity}
f_{min1} (0) = {2\sqrt{m}\over\pi\si}\,\left({n\over m} -1\right){\cal Q}(0)\,,
\end{equation}
which coincides with the expression one obtains for $f_{min1} (0)$ by 
directly substituting the large $(n,m)$ expression (\ref{Qasymptotic}) into (\ref{pmin1}), 
without first going to the scaling regime $\Delta\sim1/\sqrt{m}$, where (\ref{scalingQ})
holds.


\section{High-probability behavior}

In this paper we show that the Faybusovich vector field 
performs well with high probability, a term that is explained
in what follows.
Such an analysis was carried out for interior point methods e.g. in 
\cite{Ye-highprob,Mizuno}.
When the inputs of an algorithm have a probability distribution,
$\Delta_{\min}$ becomes a  random variable.
High probability behavior is defined as follows:
\begin{definition}
Let $T_n$ be a random variable associated with problems of size $n$.
We say that $T(n)$ is a {\em high probability bound} on $T_n$ 
if for $n \rightarrow \infty$
$T_n \leq T(n)$ with probability one.
\end{definition}

To show that $1/\Delta_{\min} < \eta(m)$ with high probability is the same
as showing that $\Delta_{\min} > 1/\eta(m)$ with high probability.
Let $f_{\min}^{(m)}(\Delta | \Delta_{\min} > 0)$ denote the probability
density of $\Delta_{\min}$ given $\Delta_{\min} > 0$.
The $m$ superscript
is a mnemonic for its dependence on the problem size.
We make the following observation:
\begin{lemma}\label{lemma-density0}
Let ${\cal P}(\Delta_{\min} > x | \Delta_{\min} > 0)$ be analytic in $x$ 
around $x=0$. Then, $\Delta_{\min} >  
\left[ f_{\min}^{(m)}(0 | \Delta_{\min} > 0) g(m) \right] ^{-1}$ 
with high-probability, 
where $g(m)$ is any function such that 
$\lim_{m \rightarrow \infty} g(m) = \infty$.
\end{lemma}

\begin{proof}
For very small $x$ we have:
\begin{equation}
{\cal P}(\Delta_{\min} > x | \Delta_{\min} > 0) \approx 1 - 
f^{(m)}_{\min}(0 | \Delta_{\min} > 0) x ~.
\end{equation}
We look for $x=x(m)$ such that 
${\cal P}(\Delta_{\min} > x(m) | \Delta_{\min} > 0) = 1$ 
with high probability.
For this it is sufficient that
\begin{equation}
 \lim_{m\rightarrow \infty} f^{(m)}_{\min}(0 | \Delta_{\min} > 0) 
x(m) = 0 \end{equation}
This holds if
\begin{equation}
x(m) =  \left[ f^{(m)}_{\min}(0 | \Delta_{\min} > 0) g(m) 
\right]^{-1} \;, \end{equation}
where $g(m)$ is any function such that 
$\lim_{m \rightarrow \infty} g(m) = \infty$.
\end{proof}

\noindent The growth of $g(m)$ can be arbitrarily slow, so 
from this point on we will ignore this factor.

\noindent As a corollary to Theorem (\ref{mainresult}) and (\ref{zerodensity})
we now obtain:
\begin{corollary} \label{corollary-deltaHighProb}
Let $(A,b,c)$ be linear programming instances distributed according to
(\ref{LPM}) then 
\beq\label{Om}
{1\over \Delta_{\min}}=\cO(m^{1/2})\,\quad\quad {\rm and}\quad\quad 
T_{\epsilon} = \cO(m^{1/2})
\eeq
with high probability.
\end{corollary}

\begin{proof}
According to the results of Section 4, (and  more explicitly, from the
derivation of (\ref{osqrtm}) in Section 7), 
$f^{(m)}_{\min}(0 | \Delta_{\min} > 0) \sim m^{1/2}$,
and the result follows from lemma (\ref{lemma-density0}) and the definition 
of $T_{\epsilon}$ (equation (\ref{tepsilon})).
\end{proof}

\begin{remark}\label{remark-tightbounds}
Note that bounds obtained in this method are tight, since
they are based on the actual distribution of the data.
\end{remark}

\begin{remark}\label{remark-nonegativemoment}
Note that $f^{(m)}_{\min}(0 | \Delta_{\min} > 0)\neq\ 0$. Therefore, the
${1\over\Delta}$ moment of the probability density function 
$f_{\min}^{(m)}(\Delta | \Delta_{\min} > 0)$
does not exist. 
\end{remark}


\section{Measures of complexity in the non-asymptotic regime}

In some situations one wants to identify the optimal vertex with limited 
precision. 

The term 
\beq\label{betadefinition}
\beta_i(t) = -\sum_{j=1}^{m} \alpha_{ji} 
\log \frac{x_{j+n-m}(t)}{x_{j+n-m}(0)}
\eeq
in (\ref{solution}), 
when it is positive, is a kind of ``barrier'':
$\Delta_{i}t$ in equation (\ref{solution}) 
must be larger than the barrier before $x_i$ can decrease to zero.

In this section we discuss heuristically the behavior of the barrier 
$\beta_i(t)$ as the dynamical system flows to the optimal vertex. 
To this end, we first discuss in rhe following sub-section some relevant 
probabilistic properties of the vertices of polyheders in our ensemble.  

\subsection{The typical magnitude of the coordinates of vertices} 
The flow (\ref{dynamics}) conserves the constraint $Ax=b$ in (\ref{standard}).
Let us split these equations according to the basic and non-basic sets which 
corresponding to an {\em arbitrary vertex} as 
\beq\label{spliteqs}
A_{\cal B}x_{\cal B} + A_{\cal N}x_{\cal N} = b\,. 
\eeq
Precisely at the vertex in question $x_{\cal N} = 0$, of course. However, we 
may be interested in the vicinity of that vertex, and thus leave $x_\cn$ 
arbitrary at this point. 

We may consider (\ref{spliteqs}) as a system of equations in the unknowns 
$x_{\cal B}$ with parameters 
$x_{\cal N}$, with coefficients $A_{\cal B}, A_{\cal N}$ and $b$ drawn from
the {\em equivariant} gaussian ensembles (\ref{aens}), (\ref{anorm}), 
(\ref{bens}) and (\ref{bnorm}). Thus, the components of $x_{\cal B}$ (e.g., 
the $x_{j+n-m}(t)$'s in (\ref{betadefinition}) if we are considering
the {\em optimal} vertex) are random variables. The joint probability 
density for the $m$ random variables $x_{\cal B}$ is given by Theorem 4.2 
of \cite{JF} (applied to the particular gaussian ensembles (\ref{aens}), 
(\ref{anorm}), (\ref{bens}) and (\ref{bnorm})) as 
\beq\label{jpdofxb}
P(x_{\cal B}; x_{\cal N}) = {\Gamma\left({m+1\over 2}\right)
\over\pi^{{m+1\over 2}}}\,{\lambda \over
\left(\lambda^2 + x_{\cal B}^T x_{\cal B}
\right)^{m+1\over 2}}\,,
\eeq
where
\beq\label{lambda}
\lambda = \sqrt{1 + x_{\cal N}^T x_{\cal N}}\,.
\eeq
(Strictly speaking, we should
constrain $x_{\cal B}$ to lie in the positive orthant, and thus multiply
(\ref{jpdofxb}) by a factor $2^m$ to keep it normalized. However, since 
these details do not affect our discussion below, we avoid introducing them
below.)

It follows from (\ref{jpdofxb}) that the components of $x_\cb$ are 
identically distributed, with probability density of any one of the 
components $x_{\cb j} = \zeta$ given by 
\beq\label{singlecomponentcauchy}
p(\zeta; x_\cn) = {1\over\pi}{\la\over\la^2 + \zeta^2}\,
\eeq
in accordance with a general theorem due to Girko \cite{Girko}.

The main object of the discussion in this sub-section is to estimate 
the typical magnitude of the $m$ components of $x_\cb$. One could argue that 
typically all $m$ components $|x_{\cb j}| <\la$, since the Cauchy distribution 
(\ref{singlecomponentcauchy}) has width $\la$. However, from 
(\ref{singlecomponentcauchy}) we have that ${\rm Prob}(|\zeta| > \la) = 1/2$, 
namely, $|x_{\cb j}| <\la$ and $|x_{\cb j}| > \la$ occur with equal 
probability. Thus, one has to be more careful, and the answer lies in 
the probability density function for $R=\sqrt{x_{\cal B}^T x_{\cal B}}$.

From (\ref{jpdofxb}), we find that the probability density function for 
$R=\sqrt{x_{\cal B}^T x_{\cal B}}$ 
takes the form  
\beq\label{probR}
\Pi\left(|x_{\cal B}| = R \right) = {2\over\sqrt{\pi}}\,
{\Gamma\left({m+1\over 2}\right)
\over\Gamma\left({m\over 2}\right)}\,{1\over \lambda}\, 
{\left({R\over\lambda}\right)^{m-1}\over 
\left[1 + \left({R\over\lambda}\right)^2\right]^{m+1\over 2}}\,.
\eeq
For a {\em finite} fixed value of $m$, this expression vanishes as 
$(R/\lambda)^{m-1}$ for $R<<\lambda$, attains its maximum at 
\beq\label{maximum}
\left({R\over\la}\right)^2 = {m-1\over 2}\,,
\eeq
and then and decays like $\lambda/R^2$ for 
$R>>\lambda$. Thus, like the even Cauchy distribution 
(\ref{singlecomponentcauchy}), it does not have a second moment.

In order to make (\ref{probR}) more transparent, we introduce the angle 
$\theta$ defined by 
\beq\label{theta}
\tan \theta\,(R) = {R\over\la}\,,
\eeq
where $0\leq\theta\leq\pi/2$. In terms of $\theta$ we have 
\beq\label{probRtheta}
\Pi\left(|x_{\cal B}| = R \right) = {2\over\sqrt{\pi}}\,
{\Gamma\left({m+1\over 2}\right)
\over\Gamma\left({m\over 2}\right)}\,{1\over \lambda}\, 
\cos^2\theta\,\sin^{m-1}\theta\,.
\eeq
(In order to obtain the probability density for $\theta$ we have to multiply 
the latter expression by a factor $dR/d\theta = \la/\cos^2\theta$.)

Let us now concentrate on the asymptotic behavior of (\ref{probRtheta}) (or 
(\ref{probR})) in the limit $m\rightarrow\infty$. Using Stirling's formula 
\beq\label{stirling}
\Gamma (x) \sim 
\sqrt{{2\pi\over x}}\,x^x\,e^{-x}
\eeq
for the large $x$ asymptotic behavior of the Gamma functions,
we obtain for $m\rightarrow\infty$
\beq\label{probRasymptotic}
\Pi\left(|x_{\cal B}| = R \right) \sim
\sqrt{2m\over\pi\la^2}\, \cos^2\theta\,\sin^{m-1}\theta\,.
\eeq
Clearly, (\ref{probRasymptotic}) is exponentially small in $m$, unless
$\sin\theta\simeq 1$, which implies 
\beq\label{thetadelta}
\theta = \pi/2 - \delta
\eeq
 with $\delta\sim
1/\sqrt{m}$. Thus, writing 
\beq\label{udelta}
\delta = \sqrt{{2u\over m}}
\eeq
(with $u<< m$), we obtain, for $m\rightarrow\infty$,
\beq\label{probRasymptoticu}
\Pi\left(|x_{\cal B}| = R \right) \sim
\sqrt{8\over\pi m\la^2}\, u\, e^{-u}\,.
\eeq
In this regime 
\beq\label{asymptoticregion}
{R\over \lambda} = \tan\theta \simeq \sqrt{m\over 2u} >> 1\,.
\eeq
The function on the r.h.s. of (\ref{probRasymptoticu}) has its maximum at 
$u=1$, i.e., at $R/\la = \sqrt{m/2}$ (in accordance with (\ref{maximum}))
and has width of ${\cal O}(1)$ around that maximum. However, this is
not enough to deduce the typical behavior of $R/\la$, since as we have 
already commented following (\ref{maximum}), 
$\Pi\left(|x_{\cal B}| = R \right)$ has long tails and decays 
like $\lambda/R^2$ past its maximum. Thus, we have to calculate the 
probability that $R>R_0 = \la\tan\theta_0$, given $R_0$. 
The calculation is straight forward: 
using (\ref{probRasymptotic}) and (\ref{theta}) we obtain
\beq\label{RbiggerR0}
{\rm Prob}(R>R_0) = \int\limits_{R_0}^\infty\,\Pi\left(R \right)\,dR = 
\sqrt{2m\over\pi}\, \int\limits_{\theta_0}^{\pi\over 2}\,
\sin^{m-1}\theta\,. 
\eeq
Due to the fact that in the limit $m\rightarrow\infty$, 
$\sin^{m-1}\theta$ may be approximated by a gaussian centered around 
$\theta = \pi/2$ with variance $1/m$, it is clear that 
$${\rm Prob}(R>R_0) = {\rm Prob}(\theta>\theta_0) \simeq 1\,,$$ unless 
$\delta_0 = \pi/2 - \theta_0 \sim \sqrt{2u_0/m}$, with $u_0<<m$. Thus, using 
(\ref{thetadelta}) and (\ref{udelta}) we obtain 
\beq\label{RbiggerR01}
{\rm Prob}(R>R_0) = \sqrt{2m\over\pi}\, \int\limits_{0}^{\delta_0}\,
\cos^{m-1}\delta \sim {1\over\sqrt{\pi}}\,\int\limits_{0}^{u_0}\,
e^{-u}\,{du\over\sqrt{u}} = {\rm erf} (\sqrt{u_0})\,.
\eeq
Finally, using the definitions of $u_0, \theta_0$ and $R_0$, we rewrite 
(\ref{RbiggerR01}) as 
\beq\label{RbiggerRfinal}
{\rm Prob}(R>R_0) = {\rm erf}\left[\sqrt{m\over 2}\,
\arctan\left({\la\over R_0}\right)\right]\,.
\eeq
From the asymptotic behavior ${\rm erf} (x) \sim 1- e^{-x^2}/x\sqrt{\pi}$
at large $x$, we see that ${\rm Prob}(R>R_0)$
saturates at 1 exponentially fast as $R_0$ decreases. Consequently, 
$1-{\rm Prob}(R>R_0)\sim \cO (m^0)$ is not negligible only if $R_0/\la$ is 
large enough, namely, $\sqrt{m\over 2}\,\arctan\left({\la\over R_0}\right) 
\leq 1$, i.e., $R_0/\la \geq \sqrt{m/2}$. If $R_0/\la$ is very large, namely,
$R_0/\la >> \sqrt{m/2}$, which corresponds 
to a small argument of the error function in (\ref{RbiggerRfinal}), 
where we clearly have ${\rm Prob}(R>R_0)\simeq \sqrt{2m/\pi}(\la/R_0) << 1$. 
From these properties of (\ref{RbiggerRfinal}) it thus follows that 
typically
\beq\label{typicalrl}
{R\over \la} \sim {\cal O}(\sqrt{m})\,.
\eeq

Up to this point, we have left the parameters $x_\cn$ unspecified. At this 
point we select the prescribed vertex of the polyheder. 
At the vertex itself, $x_{\cal N} = 0$. Therefore, from (\ref{lambda}), we 
see that $\la=1$. Thus, according to (\ref{typicalrl}), 
at the vertex, typically 
\beq\label{optimalvertex}
R_{\rm vertex}\sim \cO (\sqrt{m})\,.
\eeq 
This result obviously holds for any vertex of the polyheder: any partition 
(\ref{spliteqs}) of the system of equations $Ax=b$ into basic and non-basic 
sets leads to the same distribution function (\ref{jpdofxb}), and at each 
vertex we have $x_\cn = 0$.

Thus, clearly, this means that the whole polyheder is typically bounded 
inside an n-dimensional sphere of radius $R\sim \cO (\sqrt{m})$ centered at 
the origin.

Thus, from (\ref{optimalvertex}) and from the rotational symmetry of 
(\ref{jpdofxb}), we conclude that any component of $x_\cb$ at the optimal 
vertex, or at any other vertex (with its appropriate basic set $\cb$), is 
typically of $\cO (R_{\rm vertex}/\sqrt{m}) =  \cO (1)$ 
(and of course, positive). Points on the polyheder other than 
the vertices are weighted linear combinations of the vertices with positive 
weights which are smaller than unity, and as such 
also have their individual components typically of $\cO (1)$.

\subsection{Non-asymptotic complexity measures from $\beta_i$}

Applying the results of the previous subsection to the 
optimal vertex, we expect the components of 
$x_{\cal B}(t)$ (i.e., the $x_{j+n-m}(t)$'s in (\ref{betadefinition}))
to be typically of the same order of magnitude as their asymptotic values 
$\lim_{t\rightarrow\infty}x_\cb(t)$ at the optimal vertex, and as a result, 
we expect the barrier $\beta_i(t)$ to be of the same order of magnitude as 
its asymptotic value $\lim_{t\rightarrow\infty}\beta_i(t)$.

Note that, for this reason, in order to determine how the $x_i(t)$ in 
(\ref{solution}) tend to zero, to leading order, we can safely replace all 
the $x_{j+n-m}(t)$ by their asymptotic values in $x_{\cb}^*$.
Thus, in the following we approximate the barrier 
(\ref{betadefinition}) by its asymptotic value
\begin{equation}
\label{barrier}
\beta_i= -\lim_{t \rightarrow \infty} \sum_{j=1}^{m} \alpha_{ji} 
\log x_{j +n-m}(t) = -\sum_{j=1}^m \alpha_{ji} \log x_{j+n-m}^*\,,
\end{equation}
where we have also ignored the contribution of the initial condition.

We now consider the convergence time of the solution $x(t)$ of 
(\ref{dynamics}) to the optimal vertex. 
In order for $x(t)$ to be close to the maximum vertex we must have
$x_i(t) < \epsilon$ for $i=1,\ldots,n-m$ for some small positive 
$\epsilon$. The time parameter $t$ must then satisfy:
\begin{equation}\label{less_than_eps}
\exp (- \Delta_i t + \beta_i) < \epsilon ~,~~ \mbox{for}~ i = 1,\ldots,n-m.
\end{equation}
Solving for $t$, we find an estimate for the time required to flow to the 
vicinity of the optimal vertex as   
\begin{equation}
t > \frac{\beta_i}{\Delta_i} + \frac{|\log \epsilon |}{\Delta_i}
~, \mbox{for all}~ i = 1,\ldots,m.
\end{equation}
We define
\begin{equation}\label{T}
T = \max_{i} \left( \frac{\beta_i}{\Delta_i} + 
    \frac{|\log \epsilon |}{\Delta_i} \right)~,
\end{equation}
which we consider as the computation time.
We denote
\begin{equation}\label{betamax}
\beta_{\max} = \max_i \beta_i \;.
\end{equation}

In the limit of asymptotically small $\epsilon$, the first term in (\ref{T})
is irrelevant, and the distribution of computation times is determined by the 
distribution of the $\Delta_i$'s stated by Theorem (\ref{mainresult}).

If the asymptotic precision is not required, the first term in (\ref{T})
may be dominant. To bound this term in the expression for the computation time
we can use the quotient $\beta_{\max} / \Delta_{\min}$, where $\Delta_{min}$
is defined in (\ref{Deltamin}).

In the probabilistic ensemble used in this work $\beta_{max}$ and 
$\beta_{max}/\Delta_{min}$ are random variables, as is  $\Delta_{min}$. 
Unfortunately, we could not find the probability distributions of 
$\beta_{max}$ and $\beta_{max}/\Delta_{min}$ analytically as we did for 
$\Delta_{min}$.  In the following section, a conjecture concerning these 
distributions, based on numerical evidence, will be formulated.


\section{Scaling functions} 

In Section 4 it was shown that in the limit of large 
$(n,m)$ the probability ${\cal P}(\Delta_{min} > \Delta| \Delta_{min} > 0) $ 
is given by (\ref{mainresulteq}). Consequently, ${\cal P}(\Delta_{min} 
< \Delta|\Delta_{min} > 0) \equiv {\cal F}^{(n,m)}(\Delta)$
is of the scaling form 
\begin{equation} \label{scaling.delta}
{\cal F}^{(n,m)}(\Delta)=1-e^{x_\Delta^2}\,{\rm erfc}(x_\Delta)\ \equiv {\cal F}(x_\Delta) .
\end{equation}
Such a scaling form is very useful and informative, as we will demonstrate in 
what follows. The scaling function ${\cal F}$ contains {\em all} asymptotic 
information on $\Delta$. In particular, one can extract the problem size 
dependence of 
$f_{\min}^{(m)}(0|\Delta_{min}>0)$ which is required for obtaining a high 
probability bound using Lemma \ref{lemma-density0}. (This has already been 
shown in Corollary (\ref{corollary-deltaHighProb}).) 
We use the scaling form, equation (\ref{scaling.delta}),
leading to,
\begin{equation}\label{scaling.delta2}
f^{(m)}_{min}(0|\Delta_{min} > 0) 
= \frac{d {\cal F}^{(n,m)}(\Delta)}{d \Delta}|_{\Delta
= 0} = \eta_\Delta(n,m) \frac{{\cal F}(x_\Delta)}{dx_\Delta}|_{x_\Delta =0}.
\end{equation}
This is just $f_{min1} (0)/{\cal Q}(0)$.
With the help of lemma \ref{lemma-density0}, leading to (\ref{Om}) and our 
finding that
$\eta(n,m) \sim \sqrt{m}$, we conclude that
with high probability
\begin{equation}\label{osqrtm}
 {1\over \Delta_{\min}} = \cO(\sqrt{m}) .
\end{equation}

The next observation is that
the distribution ${\cal F}(x_\Delta)$ is very wide. For large $x_\Delta$ it behaves as 
$1-\frac{1}{\sqrt{\pi}x_\Delta}$, as is clear from the asymptotic behavior 
of the erfc function.
Therefore it does not have a mean. Since at $x_\Delta=0$ the slope $d {\cal
F}/d x_\Delta|_{x_\Delta=0}$ does not vanish, also $1/x_\Delta$ does not have a
mean (see Remark (\ref{remark-nonegativemoment})). 

We would like to derive scaling functions like (\ref{scaling.delta}) also for 
the barrier 
$\beta_{max}$, that is the maximum of the $\beta_i$ defined by (\ref{barrier}) and for the
computation time $T$ defined by (\ref{T}).
The analytic derivation of such scaling functions is difficult and therefore left for
further
studies. Their existence is verified numerically in the next section.
In particular for fixed $r=m/n$, we found that
\begin{equation} \label{scaling.beta}
{\cal P}\left(\frac{1}{\beta_{max}}<\frac{1}{\beta}\right) \equiv {\cal
F}^{(n,m)}_{1/{\beta_{max}}}\left(\frac{1}{\beta}\right)= {\cal F}_{1/{\beta}}(x_\beta)
\end{equation}
and  
\begin{equation} \label{scaling.T}
{\cal P}\left(\frac{1}{T}<\frac{1}{t}\right) \equiv {\cal
F}^{(n,m)}_{1/T}\left(\frac{1}{t}\right)= {\cal F}_{1/T}(x_T),
\end{equation}
where $\beta_{max}$ and $T$ are the maximal barrier and computation time.  
The scaling variables are 
\begin{equation}\label{scalingvbeta}
x_\beta = \eta_\beta (n,m) \frac{1}{\beta}
\end{equation}  
and
\begin{equation}\label{scalingvT}   
x_T = \eta_T (n,m) \frac{1}{t}.
\end{equation}
The asymptotic behavior of the scaling variables was determined numerically to be
\begin{equation}\label{scalingvvbeta}   
\eta_\beta (n,m) \sim m
\end{equation}
and 
\begin{equation}\label{scalingvvT}   
\eta_T (n,m) \sim m \log m.        
\end{equation}
This was found for constant $r$.
The precise $r$ dependence could not be determined numerically. 
The resulting high
probability behavior for the barrier and computation time is therefore:
\begin{eqnarray}\label{om}
\beta_{\max} = \cO(m) ,~~~
T = \cO(m \log m)
\end{eqnarray}

Note that scaling functions, such as these, immediately provide
the average behavior as well (if it exists)

Here, in the calculation of the distribution of computation times it was 
assumed that these are dominated by the barriers rather then by 
$|\log \epsilon |$ in (\ref{T}). The results (\ref{scaling.beta}), 
(\ref{scaling.T}) and (\ref{om}) are conjectures supported by the 
numerical calculations of the next section.


\section{Numerical simulations}
\label{simulations}

\begin{figure}
\epsfxsize=8cm
\centerline{\epsffile[60 200 550 610]{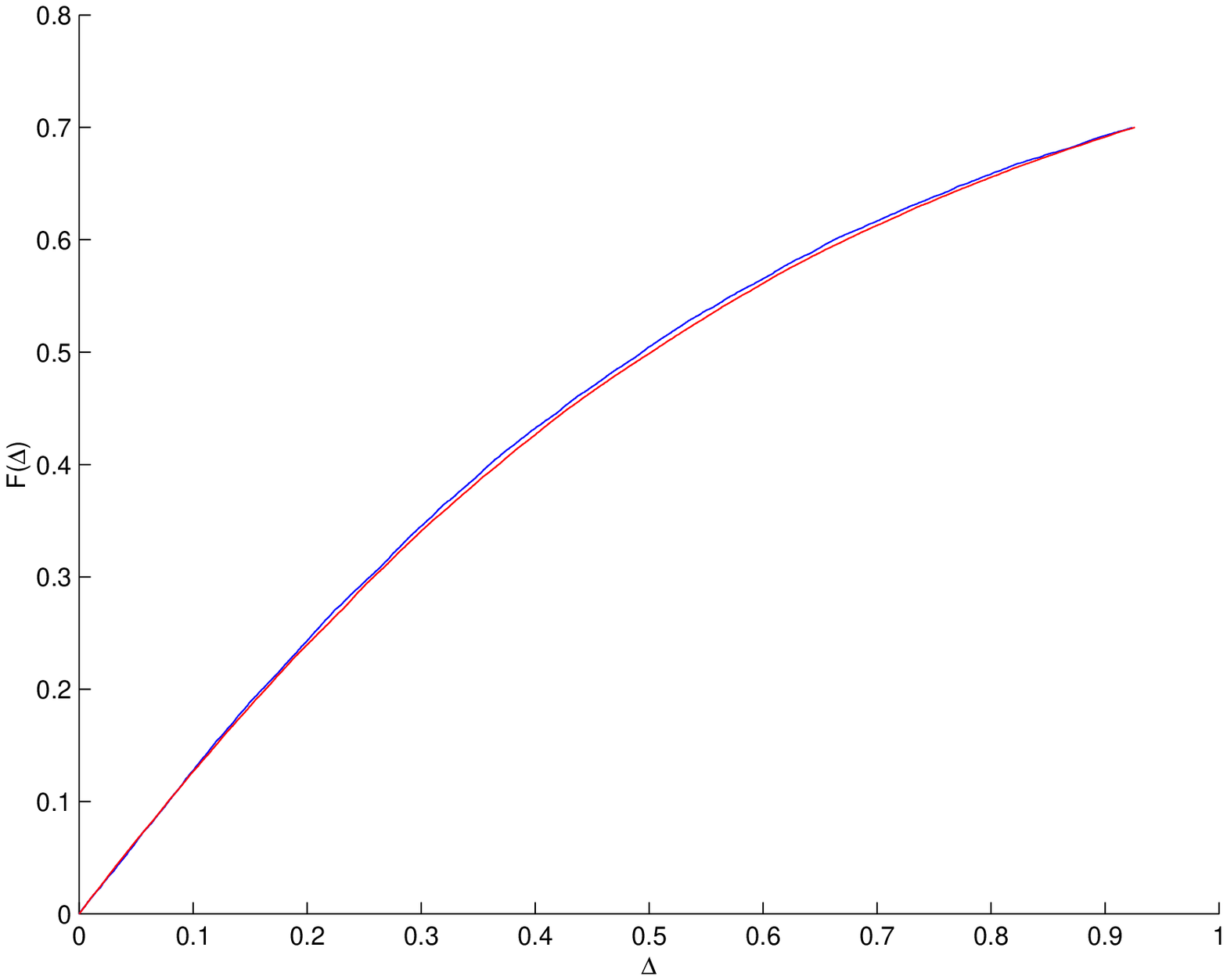}}
\caption{\label{graph-fixed_vs_opt}
Comparison of 
\ensuremath{ {\cal P}(\Delta_{min1} < \Delta | \Delta_{min1} > 0) }
and 
\ensuremath{ {\cal P}(\Delta_{\min} < \Delta | \Delta_{\min} > 0) }
for \ensuremath{m=2,n=4}.
}
\end{figure}

\begin{figure}
\epsfxsize=8cm
\centerline{\epsffile[60 200 550 610]{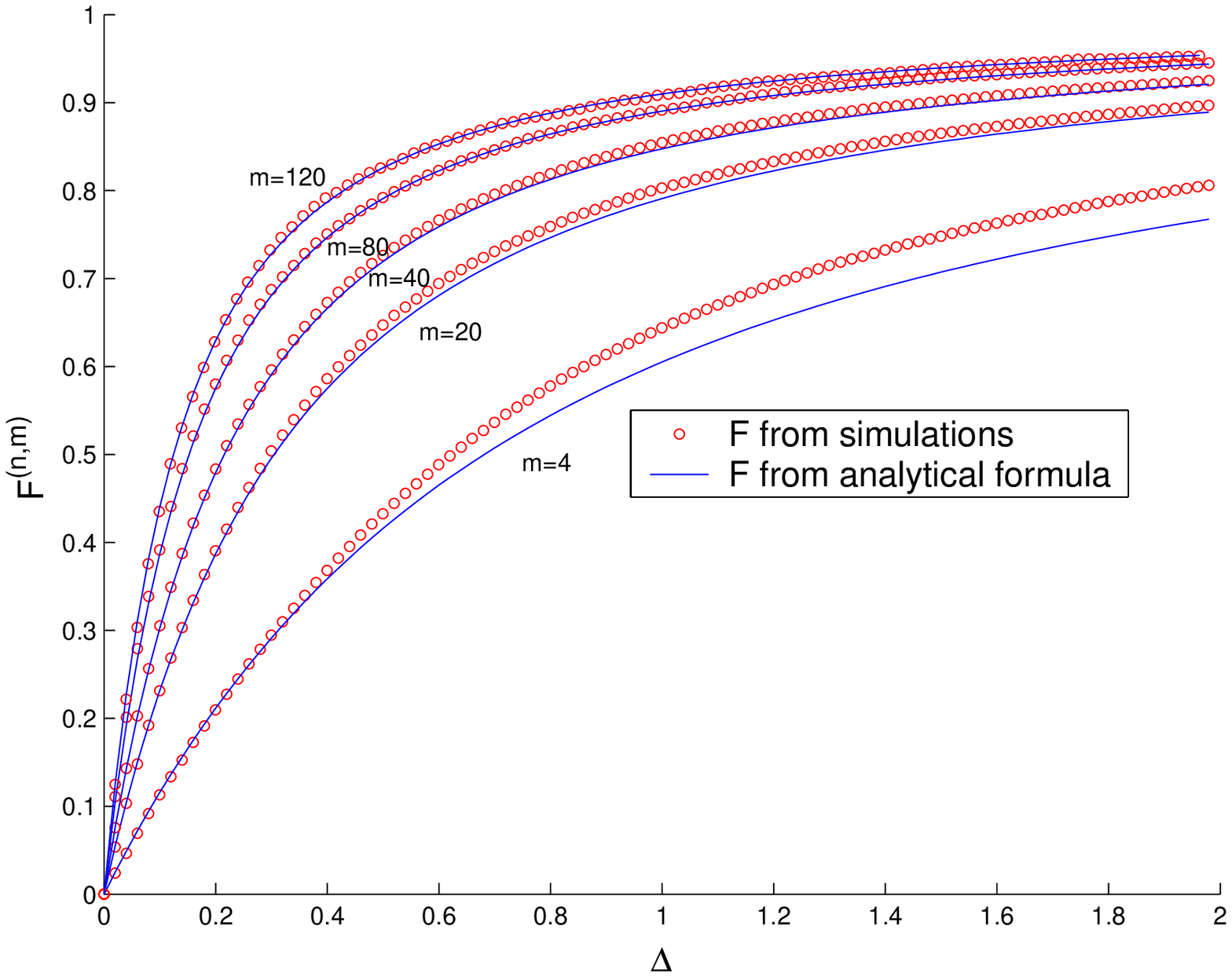}}
\caption{\label{graph-rawdelta}
\ensuremath{ {\cal F}^{(n,m)}(\Delta) }
for \ensuremath{m=4,20,40,80,120, n=2m}.
The number of instances was 
\ensuremath{10^5,10^5,40000,15000,5800 }
respectively. There is very good agreement with the
analytical results, improving as $m$ increases.
}
\end{figure}

\begin{figure}
\epsfxsize=8cm
\centerline{\epsffile[60 200 550 610]{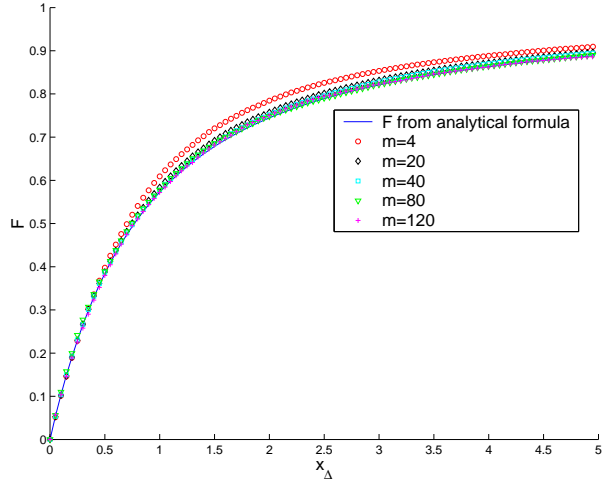}}
\caption{\label{graph-xdelta}
\ensuremath{ {\cal F}(x_\Delta) }
is plotted against the variable
\ensuremath{ x_{\Delta}},
for the same data as Figure 2.}
\end{figure}

\begin{figure}
\epsfxsize=8cm
\centerline{\epsffile[60 200 550 610]{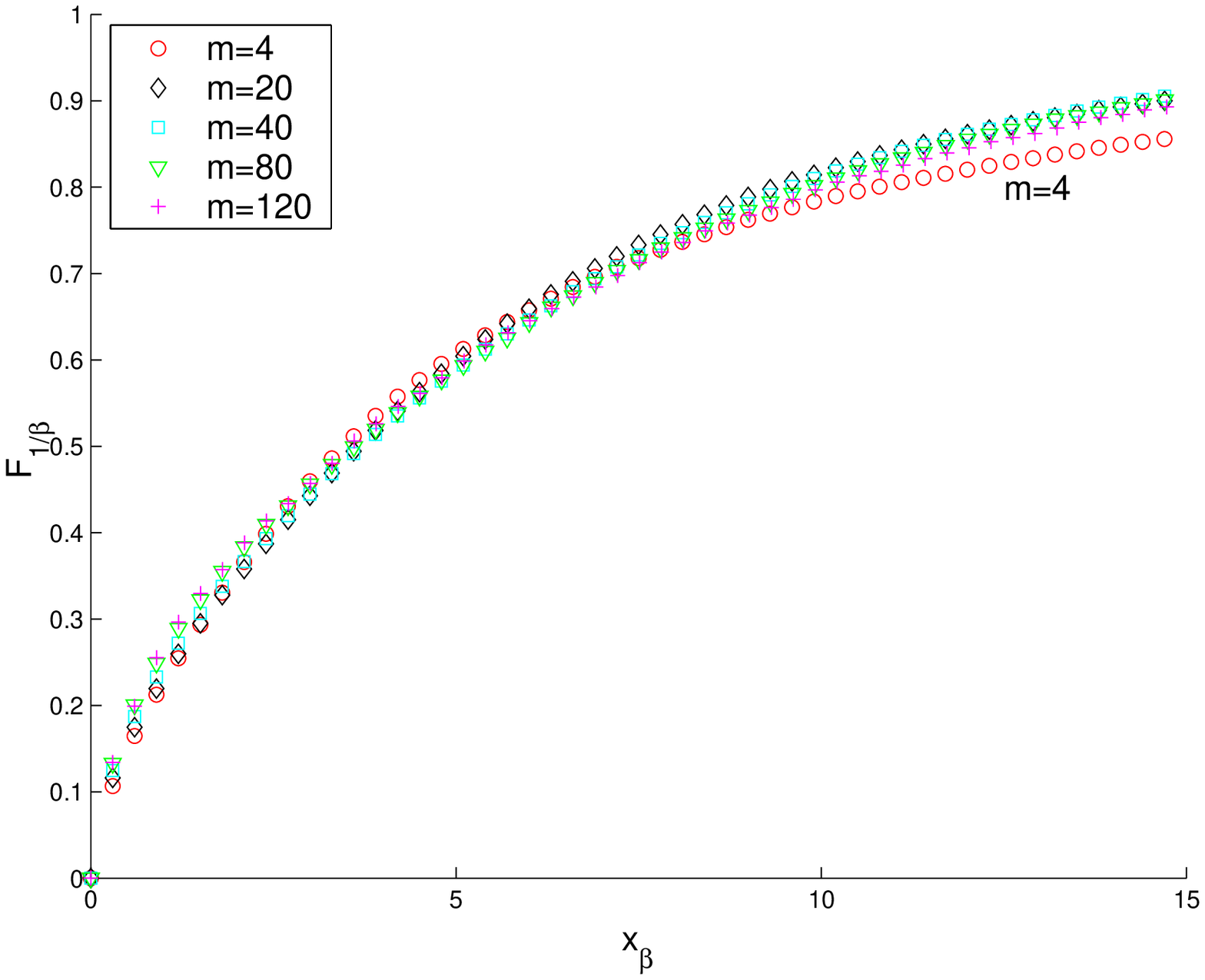}}
\caption{\label{graph-barrier}
\ensuremath{{\cal F}_{1/\beta}(x_{\beta})} as a function of the variable
\ensuremath{x_{\beta} = m / \beta_{\max} }
for the same instances as Figure 2.
}
\end{figure}

\begin{figure}
\epsfxsize=8cm
\centerline{\epsffile[60 200 550 610]{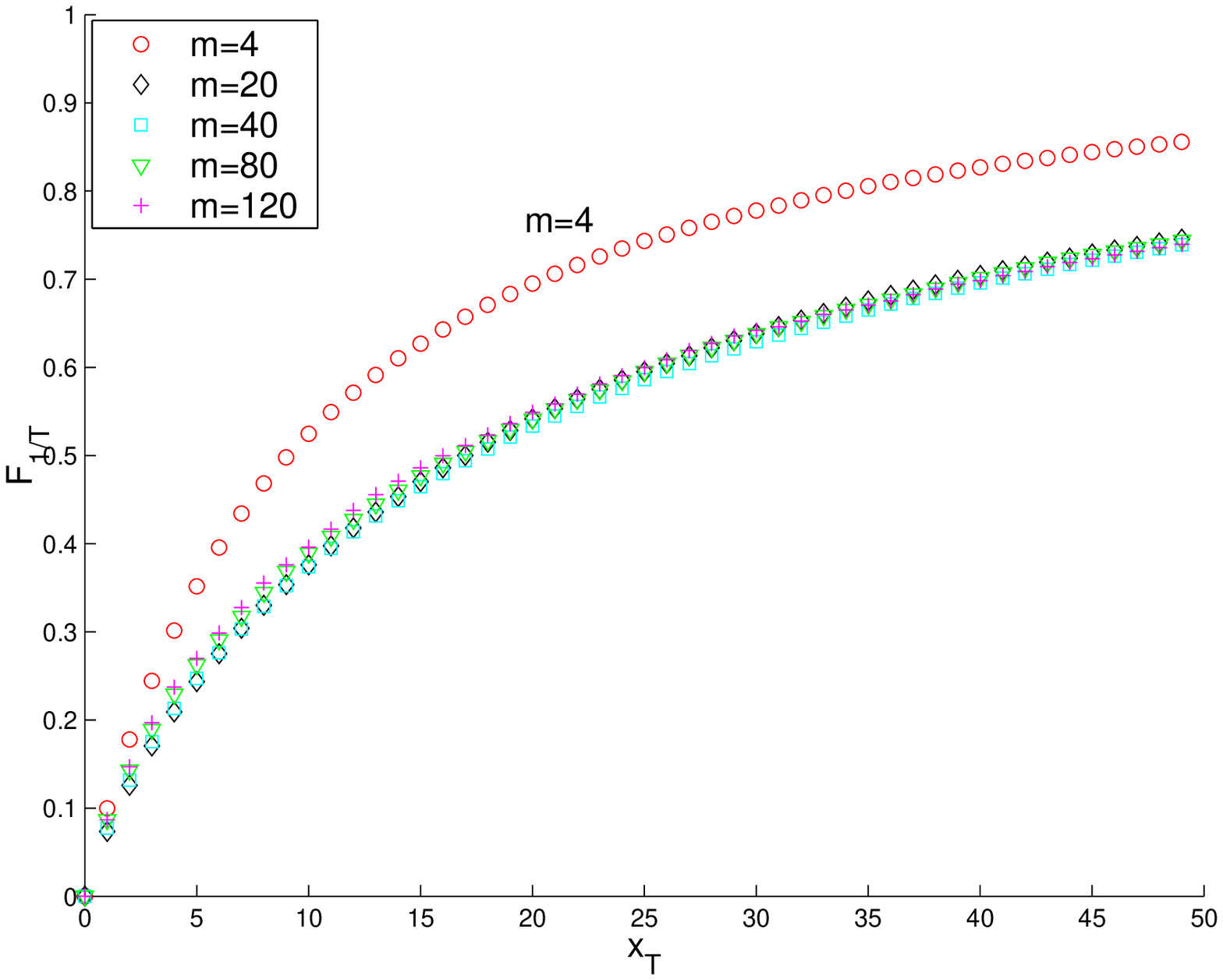}}
\caption{\label{graph-T}
\ensuremath{ {\cal F}_{1/T}(x_{T}) } as a function of the variable
\ensuremath{x_{T} = m \log m /T } for the same instances as Figure 2.
}
\end{figure}

In this section the results of numerical simulations for the distributions
of LP problems are presented.
For this purpose 
we generated full LP instances $(A,b,c)$ with the distribution
(\ref{LPM}).
For each instance the LP problem was solved using the linear programming
solver of the IMSL C library.
Only instances with a bounded optimal solution were kept, and
$\Delta_{\min}$ was computed relative to the optimal partition
and optimality was verified by checking that $\Delta_{\min} >0$.
Using the sampled instances we obtain an estimate of
${\cal F}^{(n,m)}(\Delta)={\cal P}(\Delta_{min} < \Delta |
\Delta_{min} > 0)$, and of the corresponding cumulative distribution
functions of the barrier $\beta_{\max}$ and the computation time.

As a consistency verification of the calculations we 
compared ${\cal P}(\Delta_{min} < \Delta |
\Delta_{min} > 0)$, 
to ${\cal P}(\Delta_{min1} < \Delta | \Delta_{min1} > 0)$
that was directly estimated from the distribution of matrices. 
For this purpose 
we generated a sample of $A$ and $c$ according
to the probability distributions (\ref{aens},\ref{cens}) with $\sigma=1$ and
computed
for each instance
the value of $\Delta_{min1}$ (the minimum over $\Delta_i$) for a fixed
partition of $A$ into $(N,B)$.
We kept only the positive values (note that the
definition of $\Delta_{min1}$ does not require $b$).
The two distributions are compared in Figure 1, with excellent agreement.

Note that estimation of ${\cal P}(\Delta_{min1} < \Delta | \Delta_{min1} > 0)$
by sampling from a fixed partition is infeasible for large $m$ and $n$,
since for any partition of $A$ the probability that $\Delta_{min1}$ is
positive is $2^{-(n-m)}$ (equation (\ref{Ppositive})).
Therefore the equivalence between the probability distributions of 
$\Delta_{\min}$ and $\Delta_{min1}$ cannot be exploited for producing
numerical estimates of the probability distribution of $\Delta_{\min}$.
Thus we proceed by generating full LP instances, and solving the LP
problem as described above. 

The problem size dependence was explored while keeping the ratio $n/m$ fixed or
while keeping $m$ fixed and varying $n$.
In Figure 2 we plot the numerical estimates of ${\cal F}^{(n,m)}(\Delta)$
for varying problem sizes with $n/m=2$ and compare it with
the
analytical result, Equation (\ref{scaling.delta}). 
The agreement with the analytical result improves as $m$ is increased,
since it is an asymptotic result.
The simulations show that the asymptotic result holds well even for $m=20$.
As in the analytical result, 
in the large $m$ limit we observe that ${\cal F}^{(n,m)}(\Delta)$
is not a general function of $n$, $m$ and $\Delta$, but a scaling function of
the form
${\cal F}^{(n,m)}(\Delta)={\cal F}(x_{\Delta})$ as predicted theoretically in
Section 7 (see (\ref{scaling.delta}) there). The scaling variable 
$x_{\Delta}(m)$ is given by (\ref{scalingvariable}).
Indeed, Figure 3 demonstrates that ${\cal F}^{(n,m)}$ 
has this form  as predicted by Equation
(\ref{scaling.delta}) with the scaling variable $x_\Delta$.

For the cumulative distribution functions  of the barrier $\beta_{\max}$  and 
of the computation time $T$  we do not have analytical formulas. 
These distribution functions are denoted by ${\cal F}^{(n,m)}_{1/\beta_{\max}}$ and
${\cal F}^{(n,m)}_{1/T}$ respectively. 
Their behavior near zero enables to obtain  high probability bounds on
$\beta_{\max}$ and $T$,
since for this purpose we need to bound the tails of
their
distributions, or alternatively, estimate
the density of $1/\beta{\max}$ and $1/T$ at 0. In the numerical estimate
of
the barrier we collected only positive values, since only
these contribute to prolonging the computation time.
From Figure 4 we find that ${\cal F}^{(n,m)}_{1/\beta_{\max}}$ is indeed a
scaling function of the form (\ref{scaling.beta}) with the scaling variable
$x_\beta$ of (\ref{scalingvbeta}).
The behavior of the computation time is extracted from Figure
5.
The cumulative function ${\cal F}^{(n,m)}_{1/T}$ is found to be a scaling
function
of the form (\ref{scaling.T}) with the scaling variable $x_T$ of
(\ref{scalingvT}). 
The scaling variables $x_\beta$ and $x_T$ were found numerically by the requirement
that in the asymptotic limit the cumulative distribution ${\cal F}$ approaches a
scaling form. Such a fitting is possible only if a scaling form exists.
We were unable to determine the dependence of the
scaling variables  $x_\beta$ and $x_T$ on $n/m$.

\section{Summary and discussion}

In this paper we computed the problem size dependence of the distributions of 
parameters that govern the convergence of a differential equation 
(Eq.(\ref{dynamics})) that solves the linear programming problem 
\cite{faybusovich}. To the best of our knowledge, this is the first time 
such distributions are computed. In particular, knowledge of the
distribution functions enables to obtain the high probability behavior 
(for example (\ref{osqrtm})) and (\ref{om})), and the moments 
(if these exist).

The main result of the present work is that the distribution functions of 
the convergence rate, $\Delta_{min}$, the
barrier $\beta_{max}$ and the computation time $T$ are scaling functions; 
i.e., in the asymptotic limit of large
$(n,m)$, each depends on the problem size only through a scaling variable. 
These functions are presented in section
7.

The scaling functions obtained here provide {\em all} the relevant 
information about the distribution in the
large $(n,m)$ limit.  Such functions, even if known only numerically, can be 
useful for the understanding of the
behavior for large values of $(n,m)$ that are beyond the limits of numerical 
simulations.  In particular, the distribution function of $\Delta_{min}$ was 
calculated analytically and stated as Theorem (\ref{mainresult}). The
relevance of the asymptotic theorem for finite and relatively small problem 
sizes $(n,m)$ was demonstrated numerically. It turns out to be a very simple 
function (see (\ref{scaling.delta})).  The scaling form of the
distributions of $\beta_{max}$ and of $T$ was conjectured on the basis of 
numerical simulations.

The Faybusovich flow \cite{faybusovich} that is studied in the present work, 
is defined by a system of differential
equations, and it can be considered as an example of the analysis of 
convergence to fixed points for differential equations. One should note, 
however, that the present system has a formal solution (\ref{solution}), and 
therefore it is not typical. 

If we require knowledge of the attractive fixed points with arbitrarily high 
precision (i.e., $\epsilon$ of (\ref{less_than_eps}) and (\ref{T}) can be 
made arbitrarily small), the convergence time to an $\epsilon$-vicinity of
the fixed point is dominated by the convergence rate $\Delta_{\min}$. The 
barrier, that describes the state space
``landscape'' on the way to fixed points, is irrelevant in this case. Thus, 
in this limit, the complexity is determined by (\ref{osqrtm}). This point of 
view is taken in \cite{SF}.

However, for the solution of some problems (like the one studied in the 
present work), such high precision is usually
not required, and also the non-asymptotic behavior (in $\epsilon$) of the 
vector field, as represented by the
barrier, has an important contribution to the complexity of computing the 
fixed point.

For computational models defined on the real numbers, worst case behavior 
can be ill defined and lead to infinite
computation times, in particular for interior point methods for linear 
programming \cite{BCSS}. Therefore, we compute
the distribution of computation times for a probabilistic model of 
linear programming instances rather then an upper
bound. Such probabilistic models can be useful in giving a general picture 
also for traditional discrete problem
solving, where the continuum theory can be viewed as an approximation.

A question of fundamental importance is how general is the existence of 
scaling distributions.  Their existence would
be analogous to the central limit theorem \cite{kolmogorov} and to scaling 
in critical phenomena \cite{wilson} and in
Anderson localization \cite{gangsof4, altshuler}. Typically such functions 
are universal. In the case of the central limit theorem, for example, 
under some very general conditions one obtains a Gaussian distribution, 
irrespectively of the original probability distributions. Moreover it 
depends on the random variable and the number of the original variables via
a specific combination. The Gaussian distribution is a well known example 
of the so-called stable probability distributions. 
In the physical problems mentioned above scaling and universality reflect 
the fact that the systems becomes scale invariant.

A specific challenging problem still left unsolved in the present work is the 
rigorous calculation of the distributions of $1/\beta_{max}$ and of $1/T$, 
that is proving the conjectures concerning these distributions. This
will be attempted in the near future.

\appendix

\section{The Faybusovich vector field}
\label{app-solution}

In the following we consider the inner product
$<\xi,\eta>_{X^{-1}} = \xi^{T}X^{-1} \eta$,
This inner product is defined on the positive orthant 
$\R^n_+= \{x\in \R^n : x_i > 0, \; i=1,\ldots, n \}$, where it
defines a Riemannian metric.
In the following we denote
by $a^i, \; i=1,\ldots,m$ the rows of $A$.
The Faybusovich vector field is the gradient of $h$ relative
to this metric projected to the constraint set \cite{faybusovich}.
It can be expressed as:
\begin{equation}
\mbox{grad}\, h = X c - 
\sum_{i=1}^{m} \zeta_i(x)X a^i,
\end{equation}
where $\zeta_1(x),\ldots,\zeta_m(x)$ make the gradient
perpendicular to the constraint vectors, i.e. $A \; \mbox{grad}\, h = 0$, 
so that $A x = b$ is maintained by the dynamics.
The resulting flow is 
\begin{equation}
\frac{dx}{dt}=F(x)=\mbox{grad}\, h
\end{equation}

\noindent Consider the functions
\begin{equation}
\Psi_i(x) = \log(x_i) + \sum_{j=1}^{m} \alpha_{ji} \log (x_{j+n-m})
~~i=1 \ldots n-m .
\end{equation}
The $\Psi_i$ are defined such that their equations of motion are easily
integrated.
This gives $n-m$ equations which correspond to the
$n-m$ independent variables of the LP problem.
To compute the time derivative of $\Psi_i$ we first find:
\begin{equation}
\nabla \Psi_i = \frac{1}{x_i} e^i + 
\sum_{j=1}^{m} \frac{\alpha_{ij}}{x_{j+n-m}} e^{j+n-m} \; ,
\end{equation}
and note that the vectors $\mu^i$ 
defined in equation (\ref{mu}) have the following property:
$$ < \mu^i , a^j > = 0, ~~ 
i=1,\ldots,n-m,~j =1,\ldots,m
$$
Therefore:
\begin{eqnarray}\label{Delta}
\dot{\Psi}_i(x) & = & < \nabla \Psi_i(x) , \dot{x} > =
< \nabla \Psi_i(x) , \mbox{grad}\, h >  \\ \nonumber
&=& < \mu^i , c - \sum_{j=1}^{m} \zeta_j(x) a^j> \\\nonumber
& = & < \mu^i , c> \equiv - \Delta_i   
\end{eqnarray}
This equation is integrated to yield:
\begin{equation}
x_i(t) = x_i(0) \exp \left( {-\Delta_i t - 
\sum_{j=1}^{m} \alpha_{ij} \log \frac{x_{j+n-m}(t)}{x_{j+n-m}(0)}} \right) \,.
\end{equation}


\section{The probability distribution $P(u)$}
\label{appendix-pu}

In this Appendix we study the probability distribution
function $$P(u) = \left({\lambda\over\pi}\right)^{{m^2+m\over 2}}
\int d^{m^2}B \, d^mz\, e^{-\lambda\left(\rmtr\, B^TB + z^Tz \right)}
\cdot \delta \left(u - {1\over z^T(B^TB)^{-1}z + 1}\right)\,,
$$ defined in (\ref{udistribution}) and (\ref{udistribution1})
and calculate it in detail explicitly, in the large $n,m$ limit.

We will reconstruct $P(u)$ from its moments. The $N$-th moment 
\begin{eqnarray}\label{Nmoment}
k_N &=& \int\limits_0^\infty\,du P(u)\, u^N\nonumber\\{}\nonumber\\  
&=& \left({\lambda\over\pi}\right)^{{m^2+m\over 2}}
\int d^{m^2}B \, d^mz\, e^{-\lambda\,\left(\rmtr\, B^TB + z^Tz \right)}
\left({1\over z^T(B^TB)^{-1}z + 1}\right)^N
\end{eqnarray}
of $P(u)$ may be conveniently represented as
\begin{eqnarray}\label{integralrep}
k_N &=& 
\left({\lambda\over\pi}\right)^{{m^2+m\over 2}}{1\over\Gamma(N)}\,
\int\limits_0^\infty\,t^{N-1} e^{-t}\,dt\,\int d^{m^2}B\,
e^{-\lambda\rmtr\, B^TB }\,\int\, d^mz\, e^{-z^T\left(\lambda
+{t\over B^TB}\right)z }\nonumber\\{}\nonumber\\  
&=& 
\left({\lambda\over\pi}\right)^{{m^2\over 2}}{1\over\Gamma(N)}\,
\int\limits_0^\infty\,t^{N-1} e^{-t}\,dt\,\int d^{m^2}B\,
{e^{-\lambda\rmtr\, B^TB }\over\sqrt{\det\left({\bf 1} + {t/\lambda\over B^TB}
\right)}}\,,
\end{eqnarray}
where in the last step we have performed Gaussian integration over $z$.

Recall that $P(u)$ is independent of the arbitrary parameter $\lambda$ (see
the remark preceding (\ref{udistribution1})). Thus, its $N$-th moment 
$k_N$ must also be independent of $\lambda$, which is manifest in 
(\ref{integralrep}). Therefore, with no loss of generality, and for later 
convenience, we will henceforth set $\lambda=m$ (since we have in mind
taking the large $m$ limit). Thus, 
\begin{eqnarray}\label{integralrepm}
k_N &=& 
\left({m\over\pi}\right)^{{m^2\over 2}}{1\over\Gamma(N)}\,
\int\limits_0^\infty\,t^{N-1} e^{-t}\,dt\,\int d^{m^2}B\,
{e^{-m\rmtr\, B^TB }\over\sqrt{\det\left({\bf 1} + {t/m\over B^TB}
\right)}}\nonumber\\{}\nonumber\\  
&=& 
{1\over\Gamma(N)}\,\int\limits_0^\infty\,t^{N-1} e^{-t}\,
\psi\left({t\over m}\right)\,dt\,,
\end{eqnarray}
where we have introduced the function  
\begin{equation}\label{psi}
\psi (y) = 
\left({m\over\pi}\right)^{{m^2\over 2}}\,\int d^{m^2}B\,
{e^{-m\rmtr\, B^TB }\over\sqrt{\det\left({\bf 1} + {y\over B^TB}
\right)}}\,. 
\end{equation}
Note that 
\begin{equation}\label{psinormalization}
\psi (0)=1\,.
\end{equation} 
The function $\psi (y)$ is well-defined for $y\geq 0$, where it clearly 
decreases monotonically
\beq\label{psidecrease}
\psi'(y)<0\,. 
\eeq

We would like now to integrate over the rotational degrees of freedom 
in $dB$. Any real $m\times m$ matrix $B$ may be decomposed as 
\cite{rmt, mehta}
\begin{equation}\label{decomposition}
B = {\cal O}_1^T\Omega{\cal O}_2
\end{equation}
where ${\cal O}_{1,2}\in {\cal O}(m)$, the group of $m\times m$ orthogonal 
matrices, and $\Omega = {\rm Diag}(\omega_1,\ldots ,\omega_m)$, 
where $\omega_1,\ldots,\omega_m$ are the singular values of $B$. 
Under this decomposition we may write the measure $dB$
as \cite{rmt, mehta}
\begin{equation}\label{measure}
dB = d\mu ({\cal O}_1) d\mu  ({\cal O}_2) 
\prod_{i<j} | \omega_i^2-\omega_j^2 | d^m\omega\,,
\end{equation}
where $d\mu ({\cal O}_{1,2})$ are Haar measures over the 
appropriate group manifolds. 
The measure $dB$ is manifestly invariant under actions of the orthogonal group 
${\cal O} (m)$
\begin{equation}\label{invariance}
dB=d(B{\cal O})=d({\cal O} 'B)\,,\quad\quad {\cal O}, {\cal O} ' \in {\cal O} (m)\,,
\end{equation}
as should have been expected to begin with.  

\begin{remark}
Note that the decomposition (\ref{decomposition}) is not unique, since 
$\cO_1\cd$ and $\cd\cO_2$, with $\cd$ being any of the $2^m$ diagonal matrices
${\rm Diag}\,(\pm 1, \cdots, \pm 1)$, is an equally good pair of orthogonal 
matrices to be used in (\ref{decomposition}). Thus, as $\cO_1$ and $\cO_2$ 
sweep independently over the group $\cO (m)$, the measure (\ref{measure}) 
over counts $B$ matrices. This problem can be easily rectified by appropriately
normalizing the volume ${\cal V}_m = \int d\mu ({\cal O}_1) d\mu ({\cal O}_2)$.
One can show that the correct normalization of the volume is 
\beq\label{volume}
{\cal V}_m = {\pi^{\frac{m(m+1)}{2}}\over 2^m\,\prod_{j=1}^m 
\Gamma\left(1+\frac{j}{2}\right)\Gamma\left(\frac{j}{2}\right)}\,.
\eeq
One simple way to establish (\ref{volume}), is to 
calculate $$\int dB\, \exp\,-\frac{1}{2}\rmtr B^T B = (2\pi)^{\frac{m^2}{2}} = 
{\cal V}_m\int\limits_{-\infty}^\infty d^m\omega\,
\prod_{i<j} | \omega_i^2-\omega_j^2 |\,\exp\,-\frac{1}{2}\sum_i \om_i^2\,.
$$
The last integral is a known Selberg type integral \cite{mehta}.
\end{remark}

The integrand in (\ref{psi}) depends on $B$ only through the combination 
$B^TB = {\cal O}_2^T\Omega^2{\cal O}_2$. 
Thus, the integrations over ${\cal O}_1$ and ${\cal O}_2$ in (\ref{psi}) 
factor out trivially. Thus, we end up with 
\begin{equation}\label{psi2}
\psi (y) = 
{\cal V}_m\left({m\over\pi}\right)^{{m^2\over 2}}\,
\int\limits_{-\infty}^\infty{{\prod_{i<j} | \omega_i^2-\omega_j^2 | d^m\omega} \over 
\sqrt{\det\left({\bf 1} + {y\over \Omega^2}\right)}}
\,e^{-m\rmtr\, \Omega^2}\,. 
\end{equation}
It is a straight forward exercise to check that (\ref{volume}) is consistent 
with $\psi(0)=1$.

Note that in deriving (\ref{psi2}) we have made no approximations. 
Up to this point, all our considerations in this appendix were exact. 
We are interested in the large $n,m$ asymptotic 
behavior\footnote{Recall that $m$ and $n$ tend to infinity with the ratio 
(\ref{r}), $r=m/n$, kept finite.} of $P(u)$ and of its moments. Thus, we 
will now 
evaluate the large $m$ behavior of $\psi (y)$ (which is why we 
have chosen $\lambda =m$ in (\ref{integralrepm})). This asymptotic 
behavior is determined by the saddle point dominating the integral 
over the $m$ singular values $\om_i$ in (\ref{psi2}) as $m\rightarrow\infty$.

To obtain this asymptotic behavior we rewrite the 
integrand in (\ref{psi2}) as 
$${e^{-S}\over\sqrt{\det\left({\bf 1} + {y\over \Omega^2}\right)}}\,,$$ 
where
\begin{equation}\label{gas}
S =m\,\sum_{i=1}^m \om_i^2 - {1\over 2}\,\sum_{i<j} 
\log\,(\om_i^2-\om_j^2)^2\,.
\end{equation}
In physical terms, $S$ is the energy (or the action) of the so-called 
``Dyson gas" of eigenvalues, familiar from the theory of random matrices. 

We look for a saddle point of the integral in (\ref{psi2}) in which all the 
$\om_i$ are of ${\cal O} (1)$. In such a case, $S$ in 
(\ref{gas}) is of ${\cal O} (m^2)$, and thus $e^{-S}$ overwhelms 
the factor $${1\over\sqrt{\det\left({\bf 1} + {y\over \Omega^2}\right)}} = 
e^{-{m\over 2} I(y)}\,, $$
where 
\begin{equation}\label{I}
I(y) = {1\over m}
\sum_{i=1}^m \log \left(1 + {y\over \om_i^2}\right)
\end{equation}
is a quantity of ${\cal O} (m^0) $. For later use, note that 
\begin{equation}\label{I0}
I(0) = 0\,.
\end{equation}
Thus, to leading order in $1/m$, $\psi(y)$ is dominated by the well defined 
and stable saddle point of $S$, which is indeed the case.

Simple arguments pertaining to the physics of the Dyson gas make it clear 
that the saddle point is stable: 
The ``confining potential" term $\sum_i\om_i^2 $ in (\ref{gas}) tends to 
condense all the $\om_i$ at zero, while the ``Coulomb repulsion" term 
$-\sum_{i<j}\log (\om_i^2-\om_j^2)^2$ acts to keep the $|\om_i|$ apart. 
Equilibrium must be reached as a compromise, and it must be stable, since 
the quadratic confining potential would eventually 
dominate the logarithmic repulsive interaction for $\om_i$ large enough.
The saddle point equations
\begin{equation}\label{saddle} 
{\pa S\over\pa \om_i} = 2\om_i\, 
\left[m - \sum_{j\neq i}{1\over \om_i^2-\om_j^2}\right] = 0\,, 
\end{equation}
are simply the equilibrium conditions between repulsive and attractive 
interactions, and thus determine the distribution of the $|\om_i|$.

We will solve (\ref{saddle}) (using standard techniques of random matrix 
theory), and thus will determine the equilibrium configuration of the 
molecules of the Dyson gas in the next appendix, where we show that the $m$ 
singular values $\om_i$ condense (non uniformly) into the finite 
segment (see Eq. (\ref{rhotilde})) $$ 0\leq\om_i^2\leq 2 $$
(and thus with mean spacing of the order of $1/m$).

To summarize, in the large $m$ limit, $\psi (y)$ is determined by the 
saddle point of the energy $S$ (\ref{gas}) of the Dyson gas. Thus 
for large $m$, according to (\ref{psi2}), (\ref{gas}) and (\ref{I}), 
$$\psi (y) \simeq 
{{\cal V}\over 2^m}\left({m\over\pi}\right)^{{m^2\over 2}}\,
\exp -\left(S_* + {m\over 2} I_*(y)\right)\,,$$
where $S_*$ is the extremal value of (\ref{gas}), and $I_* (y)$ is 
(\ref{I}) evaluated at that equilibrium configuration of the Dyson gas, 
namely,
\begin{equation}\label{I*}
I_*(y) = {1\over m}
\sum_{i=1}^m \log \left(1 + {y\over \om_{i*}^2}\right)\,.
\end{equation}

The actual value of $S_*$ (a number of $\cO (m^2)$) is of 
no special interest to us here, since from (\ref{psinormalization})
and (\ref{I0}) we immediately deduce that in the large $m$ limit
\begin{equation}\label{psi3}
\psi (y) \simeq e^{-{m\over 2} I_*(y)}\,.
\end{equation}

Substituting (\ref{psi3}) back into (\ref{integralrepm}) we 
thus obtain the large $(n,m)$ behavior of $k_N$ as
\beq\label{kNlargem}
k_N \simeq
{1\over\Gamma(N)}\,\int\limits_0^\infty\,t^{N-1}\,
e^{-t-{m\over 2} I_*\left({t\over m}\right)}\,dt\,.
\eeq
The function $I_*(y)$ is evaluated in the next Appendix, and is given 
in Eq. (\ref{I*final}),
$$I_*(y) = -y + \sqrt{y^2 + 2y} + \log\left(y + 1 + \sqrt{y^2 + 2y}\right)\,,
$$ which we repeated here for convenience.

The dominant contribution to the integral in (\ref{kNlargem}) comes
from values of $t<<m$, since the function 
\beq\label{exponential}
\phi (t) = t+{m\over 2} I_*\left({t\over m}\right)\,,
\eeq 
which appears in the exponent in (\ref{kNlargem}) is monotonously increasing, as can be seen from 
(\ref{positivederivative}). Thus, in this range of the variable $t$, using (\ref{smally}), we have 
\beq\label{overestimate}
\phi (t) = t + {m\over 2} I_*\left({t\over m}\right) = 
2\sqrt{2mt} + {t\over 2} 
+ \cO \left({1\over \sqrt{m}}\right)\,.
\eeq
Note that the term $t/2$ in (\ref{overestimate}) is beyond the accuracy of 
our approximation for $I_*$. The reason is that in (\ref{I**}) we used 
the continuum approximation to the density of singular values, which introduced errors 
of the orders of $1/m$. Fortunately, this term is not required. The leading order term 
in the exponential (\ref{overestimate}) of (\ref{kNlargem}) is just $\sqrt{2mt}$. Consequently, 
in the leading order (\ref{kNlargem}) reduces to 
\beqra\label{kNlargembetter}
k_N &\equiv & \int\limits_0^\infty\,du P(u)\, u^N \simeq 
{1\over\Gamma(N)}\,\int\limits_0^\infty\,t^{N-1} e^{-\sqrt{2mt}}\,dt
\nonumber\\{}\nonumber\\  
&=& {2\Gamma (2N)\over (2m)^N\Gamma(N)} = {(2N-1)!!\over m^N}\,.
\eeqra
The moments (\ref{kNlargembetter}) satisfy Carleman's criterion
\cite{bender, durrett} 
\beq\label{carleman}
\sum_{N=1}^\infty k_N^{-1/2N} = \infty\,,
\eeq
which is sufficient to guarantee that these moments define a unique 
distribution $P(u)$. 

Had we kept in (\ref{kNlargembetter}) the $\cO (m^0)$ piece of 
(\ref{overestimate}), i.e., the term $t/2$, it would have produced a 
correction factor to (\ref{kNlargembetter}) of the form $1 + \cO (N^2/m)$. 
To see this, consider the integral 
\beqast
{1\over\Gamma(N)}\,\int\limits_0^\infty\,t^{N-1} e^{-\sqrt{2mt} - {t/2}}\,dt &=& 
{2\over (2m)^N\Gamma(N)}\,\int\limits_0^\infty\,y^{2N-1} 
e^{-y - {y^2/4m}}\,dy\\{}\\ &\simeq&{2\over (2m)^N\Gamma(N)}\,\int\limits_0^\infty\,y^{2N-1} 
e^{-y}\,\left(1 - {y^2/4m} + \cdots\right)\,dy\,.
\eeqast
Thus, we can safely trust (\ref{kNlargembetter}) for moments of order $N<<\sqrt{m}$. 

The expression in (\ref{kNlargembetter}) is readily recognized 
as the $2N$-th moment of a Gaussian distribution defined on the 
positive half-line. Indeed, the moments of the Gaussian distribution 
\beq\label{gauss}
g(x;\mu) = {2\mu\over\sqrt{\pi}}\,e^{-\mu^2 x^2}\,,\quad x\geq 0
\eeq
are 
\beq\label{gaussianmoment}
\langle x^k\rangle = {\Gamma \left({k+1\over 2}\right)\over 
\sqrt{\pi}\, \mu^k}\,.
\eeq
In particular, the even moments of (\ref{gauss}) are
\beq\label{evenmoments}
\langle x^{2N}\rangle = {\Gamma \left(N+{1\over 2}\right)\over 
\sqrt{\pi}\, \mu^{2N}} = {(2N-1)!!\over (2\mu^2)^N}\,,
\eeq
which coincide with (\ref{kNlargembetter}) for $2\mu^2 = m$.
These are the moments of $u=x^2$ for the distribution $P(u)$ satisfying 
$P(u)\, du = g(x;\sqrt{m/2})dx$, as can be seen comparing (\ref{kNlargembetter}) and (\ref{evenmoments}).

Thus, we conclude that the leading asymptotic behavior of  $P(u)$ as $m$ tends to infinity
is 
\beq\label{puasymptotic}
P(u) = \sqrt{m\over 2\pi u}\,e^{-{mu\over 2}}\,,
\eeq
the result quoted in (\ref{pexplicit}).

As an additional check of this simple determination of $P(u)$ from 
(\ref{kNlargembetter}), we now sketch how to derive it more formally from the 
function 
\beq\label{resolvent}
G(z) = \int\limits_0^\infty\,{P(u)du\over z-u}\,,
\eeq
known sometimes as the Stieltjes transform of $P(u)$ \cite{bender}. 
$G(z)$ is analytic in the complex z-plane, cut along the support of $P(u)$ 
on the real axis. We can then determine $P(u)$ from (\ref{resolvent}), 
once we have an explicit expression for $G(z)$, using the identity
\beq\label{pgrelation}
P(u) = {1\over \pi}~ {\rm Im}~ G (u-i\epsilon)\,.
\eeq
For $z$ large and off the real axis, and if all the moments of $P(u)$ exist, 
we can formally expand $G(z)$ in inverse powers of $z$. Thus,
\beq\label{momentexpansion}
G(z) = \sum_{N=0}^\infty \int\limits_0^\infty\,{P(u) u^N du\over z^{N+1}}
= \sum_{N=0}^\infty {k_N\over z^{N+1}}\,.
\eeq
For the $k_N$'s given by (\ref{kNlargembetter}), the series 
(\ref{momentexpansion}) diverges. However, it is Borel summable \cite{bender}. 
Borel resummation of (\ref{kNlargembetter}), making use of 
$${1\over\sqrt{1-x}} = 1 + \sum_{N=1}^\infty \,{(2N-1)!!\over N!}\,\left({x\over 2}\right)^N\,, $$ 
yields 
\beq\label{explicitresolvent}
G(z) = {1\over z}\int\limits_0^\infty\,{e^{-t}dt\over 
\sqrt{1-{2t\over mz}}}\,.
\eeq
Thus, 
\beq\label{imaginarypart}
{1\over \pi}~ {\rm Im}~ G (u-i\epsilon) = {1\over\pi u}
\int\limits_{mu\over 2}^\infty\,{e^{-t}dt\over 
\sqrt{{2t\over mu}-1}} = \sqrt{m\over 2\pi u}\,e^{-{mu\over 2}}\,,
\eeq
which coincides with (\ref{puasymptotic}).

\pagebreak
\section{The saddle point distribution of the $\om_i$}
\label{appendix-dysongas}

We present in this Appendix the solution of the equilibrium 
condition (\ref{saddle}) of the Dyson gas of singular values 
\begin{equation}\label{saddle1}
{\pa S\over \pa\om_i^2} = m - \sum_{j\neq i}{1\over \om_i^2-\om_j^2} = 0
\eeq
(which we repeated here for convenience), and then use it to calculate 
$I_*(y)$, defined in (\ref{I*}). We follow standard 
methods \cite{bipz, rmt} of random matrix theory \cite{mehta}. 
Let 
\begin{equation}\label{som}
s_i = \om_i^2\,, 
\end{equation}
and also define 
\begin{equation}\label{gn}
F (w)={1\over m} \sum_{i=1}^{m}\langle {1\over w-
s_i}\rangle = {1\over m} \langle \rmtr {1\over w- B^TB} \rangle\,,
\end{equation}
where $w$ is a complex variable. 
Here the angular brackets denote averaging with respect to the $B$ sector 
of (\ref{aens}).
By definition, $F(w)$ behaves asymptotically as 
\begin{equation}
F(w)\wasymptotic {1\over w}\,.
\label{wasymptotic}
\end{equation}

It is clear from (\ref{gn}) that for $s>0,~ \epsilon\rightarrow 0+$ we 
have 
\beq
F (s-i\epsilon)= {1\over m} {\rm P.P.}\sum_{i=1}^m\langle {1\over 
s-s_i}\rangle + {i\pi\over m}\sum_{i=1}^m\langle\delta(s-
s_i)\rangle
\label{realim}
\eeq
where P.P. stands for the principal part. Therefore 
(from (\ref{gn})), the average eigenvalue density of $B^TB$ is given by 
\beq\label{rho}
\rho (s) \equiv {1\over m}\sum_{i=1}^m\langle\delta(s-
s_i)\rangle =  {1\over \pi}~ {\rm Im}~ F (s-
i\epsilon)\,.
\eeq
In the large $m$ limit, the real part of (\ref{realim}) 
is fixed by (\ref{saddle1}), namely, setting $s=s_i$,  
\beq
{\rm Re}~ F (s-i\epsilon) \equiv {1\over m}\langle\sum_j
{1\over s-s_j}\rangle = 1 \,.
\label{real}
\eeq

From the discussion of physical equilibrium of the Dyson gas (see the 
paragraph preceding (\ref{saddle})), we expect
the $\{s_i\}$ to be contained in  a single finite segment $0\leq s \leq a$, 
with $a$ yet to be determined. This means that $F (w)$ should have a cut 
(along the real axis, where the eigenvalues of $B^TB$ are found)
connecting $w=0$ and $a$. Furthermore, $\rho (s) $ must be integrable as 
$s\rightarrow 0+$, since a macroscopic number
(i.e., a finite fraction of $m$) of eigenvalues cannot condense at $s=0$, 
due to repulsion. These considerations, together with 
(\ref{real}) lead \cite{bipz, rmt} to the reasonable ansatz 
\beq\label{fansatz}
F (w) = 1 + \left({p\over w} + q\right)\sqrt{w(w-a)}\,,
\eeq
with parameters $p$ and $q$. 
The asymptotic behavior (\ref{wasymptotic}) then immediately fixes
\beq
q=0\,, p = - 1\,, {\rm and}~ a = 2\,.
\label{fab}
\eeq
Thus, 
\beq
F (w) = 1 - \sqrt{{w-2\over w}}\,.
\label{Gnw}
\eeq
The eigenvalue distribution of $B^TB$ is therefore
\beq\label{rhotilde}
\rho (s) = {1\over \pi}~ {\rm Im}~ F (s-i\epsilon) =  {1\over\pi }
\sqrt{{2-s\over s}}
\eeq
for $0<s<2$, and zero elsewhere.
As a simple check, note that $$\int\limits_0^2 \rho (s) ds = 1\,,$$
as guaranteed by the unit numerator in (\ref{wasymptotic}).

Thus, as mentioned in the previous appendix, $\om_i^2$, the eigenvalues of 
$B^TB$, are confined in a finite segment $0<s<2$. In the limit
$m\rightarrow\infty$, they form a continuous condensate in this segment,
with non uniform distribution (\ref{rhotilde}).

In an obvious manner, we can calculate $S_*$, the extremal value of $S$ in 
(\ref{gas}), by replacing the discrete sums over the $s_i$ by continuous 
integrals with weights $\rho (s)$ given by 
(\ref{rhotilde}). We do not calculate $S_*$ explicitly, but 
merely mention the obvious result that it is a number of ${\cal O} (m^2)$. 
Similarly, from (\ref{I*}) and (\ref{rhotilde}) we obtain
\begin{equation}\label{I**}
I_*(y) = \int\limits_0^2 \rho (s)\,\log\left( 1+ {y\over s}\right)\,ds = 
{1\over\pi }\int\limits_0^2 \,\sqrt{{2-s\over s}}
\,\log\left( 1+ {y\over s}\right)\,ds \,.
\end{equation}
Since the continuum approximation for $\rho (s)$ introduces an error
of the order $1/m$, an error of similar order is introduced in $I_*$. 
It is easier to evaluate ${dI_*(y)\over dy}$, and then integrate back, to 
obtain $I_*(y)$. We find from (\ref{I**})
\beq\label{derivative}
{dI_*(y)\over dy} = - F(-y) = -1 + {y+2\over \sqrt{y^2 + 2y}} = -1 + \sqrt{1 + 
{2\over y}}\,.
\eeq
It is clear from the last equality in (\ref{derivative}) that 
\beq\label{positivederivative}
{dI_*(y)\over dy} > 0
\eeq
for $y>0$. Integrating (\ref{derivative}), and  using (\ref{I0}), 
$I_*(0)=0$, to 
determine the integration constant, we finally obtain 
\beq\label{I*final}
I_*(y) = -y + \sqrt{y^2 + 2y} + \log\left(y + 1 + \sqrt{y^2 + 2y}\right)\,.
\eeq

From (\ref{I*final}) we obtain the limiting behaviors 
\beq\label{smally}
I_*(y) = 2\sqrt{2y} -y + \cO (y^{3/2})\,,\quad\quad 0\leq y << 1\,,
\eeq 
and 
\beq\label{yinfty}
I_*(y) = \log \left({2y\over e}\right) + 
\cO \left({1\over y}\right)\,,\quad\quad y >> 1\,.
\eeq
Due to (\ref{derivative}), $I_*(y)$ increases monotonically from 
$I_*(0)=0$ to its asymptotic form (\ref{yinfty}).
Note that  for $y=t/m$ (as required in (\ref{kNlargem})), the second
term in (\ref{smally}) is $\cO (1/m)$ and therefore it is beyond the 
accuracy of the approximation of this section.

\pagebreak


\section*{Acknowledgments}

It is our great pleasure to thank Arkadi Nemirovski,  Eduardo Sontag and Ofer
Zeitouni for stimulating and informative discussions.
This research was supported
in part by the US-Israel Binational Science Foundation
(BSF), by the Israeli Science Foundation grant number 307/98 
(090-903), by the US National Science Foundation under 
Grant No. PHY99-07949, by the Minerva Center of Nonlinear Physics 
of Complex Systems and by the fund for Promotion of Research at the Technion.


\begin{thebibliography}{10}

\bibitem{Hertz}
J.~Hertz, A.~Krogh, and R.~Palmer.
\newblock {\em Introduction to the Theory of Neural Computation}.
\newblock Addison-Wesley, Redwood City, 1991.

\bibitem{nn-optim}
A.~Cichocki and R.~Unbehauen.
\newblock {\em Neural networks for optimization and signal processing}.
\newblock John Wiley, 1993.

\bibitem{wang}
X.B. Liang and J.~Wang.
\newblock A recurrent neural network for nonlinear optimization with a
  continuously differentiable objective function and bound constraints.
\newblock {\em IEEE transaction on neural networks}, 2000.

\bibitem{mead}
C.~Mead.
\newblock {\em Analog VLSI and Neural Systems}.
\newblock Addison-Wesley, 1989.

\bibitem{brockett}
R.~W. Brockett.
\newblock Dynamical systems that sort lists, diagonalize matrices and solve
  linear programming problems.
\newblock {\em Linear Algebra and Its Applications}, 146:79--91, 1991.

\bibitem{faybusovich}
L.~Faybusovich.
\newblock Dynamical systems which solve optimization problems with linear
  constraints.
\newblock {\em IMA Journal of Mathematical Control and Information},
  8:135--149, 1991.

\bibitem{helmke-moore}
U.~Helmke and J.B. Moore.
\newblock {\em Optimization and Dynamical Systems}.
\newblock Springer Verlag, London, 1994.

\bibitem{Branickyshort}
M.S. Branicky.
\newblock Analog computation with continuous {ODE}s.
\newblock In {\em Proceedings of the IEEE Workshop on Physics and Computation},
  pages 265--274, Dallas, TX, 1994.

\bibitem{Papadimitriou}
C.~Papadimitriou.
\newblock {\em Computational Complexity}.
\newblock Addison-Wesley, Reading, Mass., 1995.

\bibitem{realtime}
L.O. Chua and G.N. Lin.
\newblock Nonlinear programming without computation.
\newblock {\em IEEE transaction on circuits and systems}, 31(2), 1984.

\bibitem{dds2}
A.~Ben-Hur, H.T. Siegelmann, and S.~Fishman.
\newblock A theory of complexity for continuous time dynamics.
\newblock Accepted, Journal of Complexity.

\bibitem{SF}
H.T. Siegelmann and S.~Fishman.
\newblock Computation by dynamical systems.
\newblock {\em Physica D}, 120:214--235, 1998.

\bibitem{BCSS}
L.~Blum, F.~Cucker, M.~Shub, and S.~Smale.
\newblock {\em Complexity and real Computation}.
\newblock Springer-Verlag, 1999.

\bibitem{lptraub}
J.F. Traub and H.~Wozniakowski.
\newblock Complexity of linear programming.
\newblock {\em Operations Research Letters}, 1:59--62, 1982.

\bibitem{renegar-condition}
J.~Renegar.
\newblock Incorporating condition measures into the complexity theory of linear
  programming.
\newblock {\em SIAM J. Optimization}, 5(3):506--524, 1995.

\bibitem{lpsmale}
S.~Smale.
\newblock On the average number of steps in the simplex method of linear
  programming.
\newblock {\em Math. Programming}, 27:241--262, 1983.

\bibitem{Todd-models}
M.J. Todd.
\newblock Probabilistic models for linear programming.
\newblock {\em Mathematics of Operations Research}, 16:671--693, 1991.

\bibitem{shamir}
R.~Shamir.
\newblock The efficiency of the simplex method: A survey.
\newblock {\em Management Science}, 33(3):301--334, 1987.

\bibitem{Anstreicher}
K.M. Anstreicher, J.~Ji, F.A. Potra, and Y.~Ye.
\newblock Probabilistic analysis of an infeasible interior-point algorithm for
  linear programming.
\newblock {\em Mathematics of Operations Research}, 24:176--192, 1999.

\bibitem{Ye-book}
Y.~Ye.
\newblock {\em Interior Point Algorithms: Theory and Analysis}.
\newblock John Wiley and Sons Inc., 1997.

\bibitem{Ye-highprob}
Y.~Ye.
\newblock Toward probabilistic analysis of interior-point algorithms for linear
  programming.
\newblock {\em Mathematics of Operations Research}, 19:38--52, 1994.


\bibitem{Saigal}
R. Saigal.
{\em Linear Programming}.
Kluwer Academic, 1995.


\bibitem{bipz}
E. Br\'{e}zin, C. Itzykson, G. Parisi and J.~-B. Zuber, Planar diagrams. {\it
Comm.~Math.~Phys.} {\bf 59} (1978) 35. 


\bibitem{rmt}
Some papers that treat random {\em real} rectangular matrices, such as the
matrices relevant for this work, are:\\
A.~ Anderson, R.~C. Myers and V.~ Periwal, Complex random surfaces. 
{\it Phys. Lett.}{\bf B 254}(1991) 89,\\
Branched polymers from a double scaling limit of matrix models.
 ~~ {\it Nucl. ~Phys.} {\bf B 360}, (1991) 463 (Section 3).\\
J.~Feinberg and A.~Zee, Renormalizing rectangles and other topics in random 
matrix theory. {\it J. Stat. Mech.} {\bf 87} 
(1997) 473-504.\\
For earlier work see: G.M. Cicuta, L. Molinari, E. Montaldi and F. Riva,  
Large rectangular random matrices.
{\it J. Math.Phys.} {\bf 28} (1987) 1716.  

\bibitem{mehta}
M.L. Mehta.
\newblock {\em Random Matrices}.
\newblock Academic Press, Boston, 2nd edition edition, 1991.

\bibitem{Mizuno}
S.~Mizuno, M.J. Todd, , and Y.~Ye.
\newblock On adaptive-step primal-dual interior-point algorithms for linear
  programming.
\newblock {\em Mathematics of Operations Research}, 18:964--981, 1993.

\bibitem{JF}
J. Feinberg. 
\newblock On the universality of the probability distribution of 
the product $B^{-1}X$ of random matrices. 
\newblock arXiv:math.PR/0204312, 2002. 

\bibitem{Girko}
V.L. Girko.
\newblock On the distribution of solutions of systems of linear equations with
  random coefficients.
\newblock {\em Theory of probability and mathematical statistics}, 2:41--44,
  1974.

\bibitem{kolmogorov}
B.V. Gnedenko and A.N. Kolmogorov, 
\newblock {\em Limit Distributions for Sums of Independent Random
Variables}.
\newblock Addison Wesley, Reading, MA, 1954.

\bibitem{wilson}
K.G. Wilson and J. Kogut,
\newblock {\em The renormalization group and the epsilon expansion}.
{\it Phys.~Rep.} {\bf 12}, (1974) 75.\\
J. Cardy, 
\newblock {\em Scaling and Renormalization in Statistical Physics}.
\newblock Cambridge University Press, Cambridge, 1996.


\bibitem{gangsof4}
E. Abrahams, P. W. Anderson, D. C. Licciardelo and T. V. Ramakrishnan, 
Scaling theory of localization: absence of quantum diffusion in 
two dimensions. {\it Phys. ~Rev. ~Lett.} {\bf 42} (1979) 673;\\
E. Abrahams, P. W. Anderson, D. S. Fisher and D. J. Thouless, 
New method for a scaling theory of localization. 
{\it Phys. ~Rev.} {\bf B 22} (1980) 3519.\\
P. W. Anderson, New method for scaling theory of localization. II. 
Multichannel theory of a "wire" and possible extension to higher
dimensionality. {\it Phys. ~Rev.} {\bf B 23} (1981) 4828.


\bibitem{altshuler}
B.L. Altshuler, V.E. Kravtsov and I.V. Lerner, in 
\newblock {\em Mesoscopic Phenomena in Solids}, ed. B.L. Altshuler, P.A. Lee 
and R.A. Webb.
\newblock North Holland, Amsterdam, 1991.

\bibitem{bender}
C.M. Bender and S.A. Orszag. 
\newblock {\em Advanced Mathematical Methods for Scientists and Engineers}.
\newblock Springer Verlag, New York, 2nd edition, 1999. (Chapter 8).


\bibitem{durrett}
R. Durrett. 
\newblock {\em Probability: Theory and Examples}.
\newblock Wadswarth Publishing Co., Belmont, 2nd edition, 1996. (Chapter 2).

\end{thebibliography}
\end{document}